\documentclass[aps,twocolumn,preprintnumbers,amsmath,amssymb,superscriptaddress]{revtex4-1} 

\usepackage[bottom]{footmisc}

\usepackage{graphicx}
\usepackage{dcolumn}
\usepackage{color}
\usepackage{amsmath}
\usepackage[breaklinks,colorlinks,bookmarks=false,citecolor=blue,linkcolor=red,urlcolor=blue]{hyperref}

\usepackage{tikz}

\newcommand{\ket}[1]{\ensuremath{\left|#1\right\rangle}}

\newcommand{\angstrom}{\mbox{\normalfont\AA}}

\newcommand\shastrysutherlandlatticefullapproachpCUTEzero[1][t]{
\begin{tikzpicture}[darkstyle/.style={circle,draw,fill=black,minimum size=15}]

\foreach \x in {0,...,5}
    \foreach \y in {0,...,5} 
       \node [white]  (\x\y) at (2*\x,2*\y) { };
  \foreach \x in {1,...,4}
    \foreach \y [count=\yi] in {0,...,4}  
  	  \pgfmathtruncatemacro{\cur}{\y +1}
      \path (\x\y) edge [black] node {} (\x\cur)
      		(\y\x)edge [black] node {} (\yi\x);

\foreach \y in {1,3}
    \foreach \x [count=\xi] in {1,3}
  	  \pgfmathtruncatemacro{\xr}{\x +1}
  	  \pgfmathtruncatemacro{\yo}{\y +1}
	  \draw [fill=gray!50,opacity=.5] (\x\y) rectangle (\xr\yo);

\foreach \x in {1,...,4}
    \foreach \y in {1,...,4} 
       \node [darkstyle]  (\x\y) at (2*\x,2*\y) { };
  \foreach \x in {1,...,4}
    \foreach \y [count=\yi] in {1,...,3}  
      \path (\x\y) edge [black, dotted] node {} (\x\yi)
      		(\y\x)edge [black, dotted] node {} (\yi\x);
 \foreach \x in {1,...,4}
    \foreach \y [count=\yi] in {1,3}
  	  \pgfmathtruncatemacro{\cur}{\y +1}
      \path (\x\y) edge [black, line width=6pt] node {} (\x \cur);
 \foreach \y in {1,...,4}
    \foreach \x [count=\xi] in {1,3}
  	  \pgfmathtruncatemacro{\cur}{\x +1}
      \path (\x\y) edge [black, line width=6pt] node {} (\cur \y);   
      
\foreach \y in {0,2,4}
    \foreach \x [count=\xi] in {0,2,4}
  	  \pgfmathtruncatemacro{\xn}{\x +1}
  	  \pgfmathtruncatemacro{\yn}{\y +1}
      \path (\x\y) edge [dashed, very thick] node {} (\xn \yn);
\foreach \y in {1,3}
    \foreach \x [count=\xi] in {2,4}
  	  \pgfmathtruncatemacro{\xn}{\x -1}
  	  \pgfmathtruncatemacro{\yn}{\y +1}
      \path (\x\y) edge [dashed, line width=3pt] node {} (\xn \yn);   
      
\def\size{2.1}
\def\sizeone{2.}
\node[black] (nJone) at (3,8.45) {\scalebox{\size}{$1$}};
\node[black] (nJtwo) at (5,7.6) {\scalebox{\size}{$\lambda_{\text{F}} J_2'$}};
\node[black] (nJ) at (3.15,7.35) {\scalebox{1.5}{$J_1^0$}};
\node[black] (nJ) at (2.9,6.3) {\scalebox{1.3}{$\lambda_{\text{F}} \Delta J_1$}};
\node[black] (nJsecond) at (5.2,4.5) {\scalebox{1.5}{$\lambda_{\text{F}} J_2$}};
\end{tikzpicture}
}

\newcommand\shastrysutherlandlatticeemptyapproachpCUT[1][t]{
\begin{tikzpicture}[darkstyle/.style={circle,draw,fill=black,minimum size=15}]

\foreach \x in {0,...,5}
    \foreach \y in {0,...,5} 
       \node [white]  (\x\y) at (2*\x,2*\y) { };
  \foreach \x in {1,...,4}
    \foreach \y [count=\yi] in {0,...,4}  
  	  \pgfmathtruncatemacro{\cur}{\y +1}
      \path (\x\y) edge [black] node {} (\x\cur)
      		(\y\x)edge [black] node {} (\yi\x);

\foreach \y in {1,3}
    \foreach \x [count=\xi] in {1,3}
  	  \pgfmathtruncatemacro{\xr}{\x +1}
  	  \pgfmathtruncatemacro{\yo}{\y +1}
	  \draw [fill=gray!50,opacity=.5] (\x\y) rectangle (\xr\yo);

\foreach \x in {1,...,4}
    \foreach \y in {1,...,4} 
       \node [darkstyle]  (\x\y) at (2*\x,2*\y) { };
 \foreach \x in {1,...,4}
    \foreach \y [count=\yi] in {1,3}
  	  \pgfmathtruncatemacro{\cur}{\y +1}
      \path (\x\y) edge [black, line width=6pt] node {} (\x \cur);
 \foreach \y in {1,...,4}
    \foreach \x [count=\xi] in {0,2,4}
  	  \pgfmathtruncatemacro{\cur}{\x +1}
      \path (\x\y) edge [black, line width=6pt] node {} (\cur \y);   
      
\foreach \y in {0,2,4}
    \foreach \x [count=\xi] in {1,3}
  	  \pgfmathtruncatemacro{\xn}{\x +1}
  	  \pgfmathtruncatemacro{\yn}{\y +1}
      \path (\x\y) edge [dashed, very thick] node {} (\xn \yn);
\foreach \y in {1,3}
    \foreach \x [count=\xi] in {1,3,5}
  	  \pgfmathtruncatemacro{\xn}{\x -1}
  	  \pgfmathtruncatemacro{\yn}{\y +1}
      \path (\x\y) edge [dashed, line width=3pt] node {} (\xn \yn);   

\def\size{2.1}
\def\sizeone{2.}
\node[black] (nJone) at (2.5,6.9) {\scalebox{\size}{$1$}};
\node[black] (nJofe) at (1.05,8.5) {\scalebox{\size}{$\lambda_{\text{E}}$}};
\node[black] (nJtwo) at (3.4,8.65) {\scalebox{1.5}{$\lambda_{\text{E}} J_2$}};
\node[black] (nJsone) at (4.7,6.65) {\scalebox{1.5}{$\lambda_{\text{E}} J_1$}};
\node[black] (nJone) at (1.2,5) {\scalebox{\size}{$\lambda_{\text{E}} J_2'$}};
\node[black] (nJond) at (7,6.3) {\scalebox{\sizeone}{$\tau_{\text{horiz}}$}};
\end{tikzpicture}
}

\begin{document}

\title{Competition between intermediate plaquette phases in SrCu$_2$(BO$_3$)$_2$ under pressure}

\author{C.~Boos}
\email{carolin.boos@fau.de}
\affiliation{Institute for Theoretical Physics, FAU Erlangen-N\"urnberg, Germany}
\affiliation{Institute of  Physics, Ecole Polytechnique F\'{e}d\'{e}rale de Lausanne (EPFL), CH 1015 Lausanne, Switzerland}

\author{S.P.G.~Crone}
\affiliation{Institute for Theoretical Physics and Delta Institute for 
Theoretical Physics, University of Amsterdam, Science Park 904, 1098 XH 
Amsterdam, The Netherlands}

\author{I.A.~Niesen}
\affiliation{Institute for Theoretical Physics and Delta Institute for 
Theoretical Physics, University of Amsterdam, Science Park 904, 1098 XH 
Amsterdam, The Netherlands}

\author{P.~Corboz}
\affiliation{Institute for Theoretical Physics and Delta Institute for 
Theoretical Physics, University of Amsterdam, Science Park 904, 1098 XH 
Amsterdam, The Netherlands}

\author{K.~P.~Schmidt}
\email{kai.phillip.schmidt@fau.de}
\affiliation{Institute for Theoretical Physics, FAU Erlangen-N\"urnberg, Germany}

\author{F.~Mila}
\email{frederic.mila@epfl.ch}
\affiliation{Institute of Physics, Ecole Polytechnique F\'{e}d\'{e}rale de Lausanne (EPFL), CH 1015 Lausanne, Switzerland}

\date{\today}

\begin{abstract}
Building on the growing evidence based on NMR, magnetization, neutron scattering, ESR, and specific heat that, under pressure, SrCu$_2$(BO$_3$)$_2$ has an intermediate phase between the dimer and the N\'eel phase, we study the competition between two candidate phases in the context of a minimal model that includes two types of intra- and inter-dimer interactions without enlarging the unit cell. We show that the empty plaquette phase of the Shastry-Sutherland model is quickly replaced by a quasi-1D full plaquette phase when intra- and/or inter-dimer couplings take different values, and that this full plaquette phase is in much better agreement with available experimental data than the empty plaquette one.
\end{abstract}

\maketitle

Almost two decades after the discovery of the first magnetization plateaus, the investigation of the layered material SrCu$_2$(BO$_3$)$_2$ under extreme conditions continues to attract a lot of attention and to reveal new fascinating properties. If there is by now ample evidence in favor of a sequence of magnetization plateaus at 1/8, 2/15, 1/6, 1/4, 1/3, and 1/2 (and possibly 2/5)~\cite{Kageyama99,Onizuka00,Kodama02,takigawa04,levy08,Sebastian08,Jaime12,takigawa13,matsuda13,haravifard16},  the structure of some of these plateaus remains debated, and several groups are attempting to perform X-ray or neutron scattering in fields above 27 Tesla and at very low temperature to have direct information on the structure of the 1/8 plateau. In parallel, the investigation of the phase diagram under pressure using various techniques ranging from NMR~\cite{doi:10.1143/JPSJ.76.073710} to magnetization~\cite{haravifard16}, ESR~\cite{doi:10.7566/JPSJ.87.033701}, neutron scattering~ \cite{Zayed_plaquette_16}, and specific heat~\cite{sandvik2019} has revealed the presence of a phase transition at around 1.7 GPa to a new gapped phase that is the subject of the present paper. 

%
\begin{figure}[t]
\begin{center}
\includegraphics[trim={5cm 19.7cm 6.9cm 2cm}, clip=true, width=0.45\textwidth]{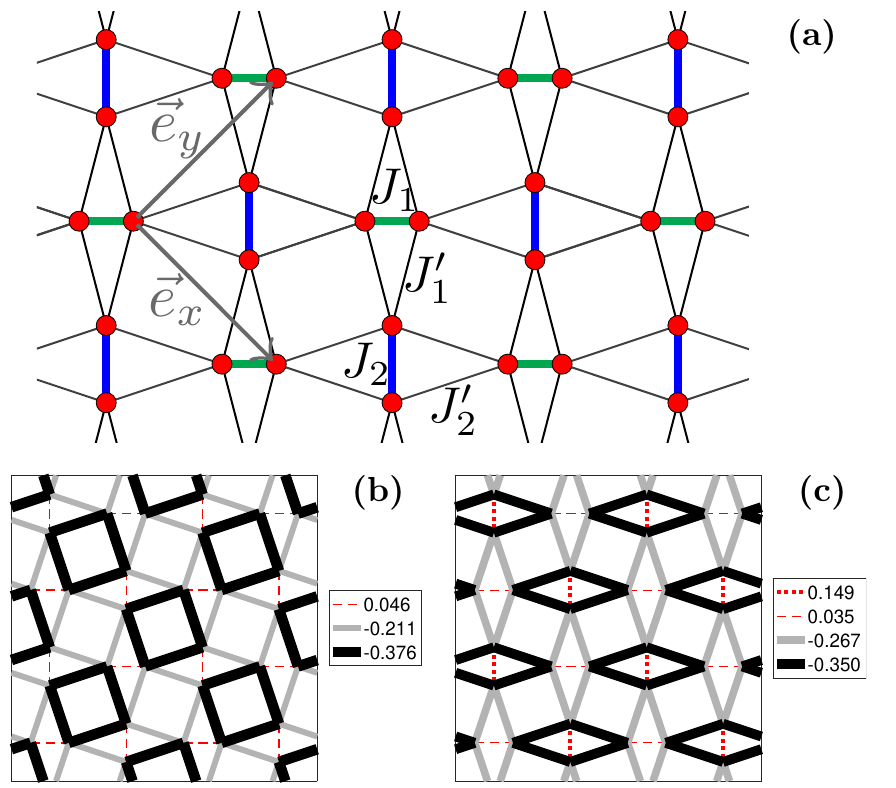}
\caption{(a) Sketch of the distorted orthogonal dimer lattice. 
The unit cell is defined by the unit vectors $\vec{e}_x$ and $\vec{e}_y$.
(b) Spin-spin correlations in the EPP and (c) in the FPP phase obtained with iPEPS for the  parameters used in Fig.~3a,b and 3d,e, respectively.}
\label{Fig1}
\end{center}
\end{figure}

SrCu$_2$(BO$_3$)$_2$ is described to a very good accuracy by the stacking of the 2D Shastry-Sutherland model~\cite{ShaSu81}, also known as the orthogonal dimer model~\cite{MiUeda03}, defined by the Hamiltonian
\begin{equation}
H = J \sum_{\langle\langle i,j \rangle\rangle} \vec S_i \cdot \vec S_j + J' \, \sum_{ 
\langle i,j \rangle} \,\vec S_i \cdot \vec S_j,
\label{ess}
\end{equation}
where $J$ is the intra-dimer coupling and $J'$ the inter-dimer coupling. In the limit $J'=0$, the system consists of a set of decoupled dimers, and the exact ground state is a product of singlets on these dimers. Due to the frustrated nature of the inter-dimer coupling, this remains strictly true as long as $J'$ is not too large. In the opposite limit $J=0$, the system is a square lattice with nearest-neighbor antiferromagnetic couplings, and the ground state possesses long-range N\'eel order. In between, there is an intermediate phase that, after some debate~\cite{Albrecht96,Miyahara99,Zheng99,Muller00,koga00,Takushima01,Chung01,Laeuchli02,PhysRevB.65.014408}, has been convincingly proven to be the empty plaquette phase (EPP) depicted in Fig.~\ref{Fig1}b and to exist in the range $0.675(2)<J'/J<0.765(15)$~\cite{PhysRevB.87.115144}. The dominant interlayer interactions are not expected to change the physics qualitatively since the product of dimer singlets is still a ground state, while the small Dzyaloshinskii-Moriya interactions are not expected to shift the boundaries significantly since their effect on the ground-state energy of a gapped singlet phase is of second order.

Since up to an overall energy scale the ratio $J'/J$ is the only parameter of the Shastry-Sutherland model, applying hydrostatic pressure to change this ratio is a natural way to probe this phase diagram, and this has been first attempted in 2007 using NMR~\cite{doi:10.1143/JPSJ.76.073710}. This experiment has indeed revealed the presence of a new phase at 2.4 GPa, but in this intermediate phase, there are two types of Cu sites. This is incompatible with the EPP, in which all Cu sites remain equivalent. The report of a weak orthorhombic distortion already at low pressure has led to the investigation of a model with two sets of intra-dimer couplings~\cite{PhysRevB.83.140414}. If the couplings are sufficiently different, another intermediate phase is realized. It is a one-dimensional phase related to a spin-one Haldane chain. Note however that the presence of an orthorhombic distortion at low pressure has not been confirmed by subsequent experiments.

More recently, neutron scattering experiments have confirmed the presence of 
an intermediate phase~\cite{Zayed_plaquette_16} characterized by the presence of an additional second excitation branch at low energies, in sharp contrast with the dimer phase. The structure factors of these excitations appear to be incompatible with the EPP, another indication that the intermediate phase is not that of the Shastry-Sutherland model. They are however compatible with a putative full plaquette phase (FPP) in which bonds get stronger around plaquettes with diagonal couplings (see Fig.~\ref{Fig1}c). 

In this Letter, we discuss theoretically the possible nature of this intermediate phase. We show that the Haldane and the FPP actually build a single phase in the phase diagram of a generalized Shastry-Sutherland model that includes two types of intra and inter-dimer couplings, and that the properties of this phase are in much better agreement with available experimental data than those of the EPP. Consequences for the three dimensional system are briefly discussed.

In choosing the model to describe the competition between different possible intermediate phases, we have paid special attention to the fact that, so far, no distortion could be detected (although there has to be one of course, as discussed at the end of the paper). So we have concentrated on a minimal modification that contains two sets of inequivalent $J$-bonds, $J_1$ and $J_2$, as assumed in Ref.~\cite{PhysRevB.83.140414}, but also two sets of inequivalent $J'$-bonds, $J_1'$ and $J_2'$ (see Fig.~\ref{Fig1}a). Models with inequivalent $J'$-bonds have been introduced in Ref.~\onlinecite{Takushima01} as starting points of series expansions (SE), but the relative stability of the EPP and FPP phases has not been studied. The first goal of the present paper is to map out precisely these stability regions. As pointed out recently by Lee et al~\cite{sachdev2019}, the two candidate plaquette phases correspond to natural distortions in a Landau expansion, depending on the sign of the coupling constant. In the FPP, diamonds with short intra- and inter-dimer bonds form. Naively one could expect both intra- and inter-dimer couplings to get stronger, but this is not the case. The intra-dimer coupling corresponds to a Cu-O-Cu bond with an angle of 97.6 degrees, and making it shorter will actually {\it decrease} the magnitude of the coupling constant~\cite{doi:10.7566/JPSJ.87.033701}. By contrast, the inter-dimer coupling is a more standard geometry, and the coupling constant is expected to get stronger if the bond gets shorter. So we have considered the parameter range where the weaker intra-dimer coupling $J_1$ is surrounded by stronger inter-dimer couplings $J'_1$. Taking the other configuration would anyway require very different inter-dimer bonds to stabilize the FPP, a possibility which is not realistic.

Our results have been obtained by two complementary methods, infinite projected entangled-pair states (iPEPS) and high-order series expansions (SE). An iPEPS is a  variational tensor-network ansatz for two-dimensional ground states in the thermodynamic limit~\cite{verstraete_renormalization_2004,nishio2004,jordan_classical_2008}, where the accuracy is systematically controlled by the bond dimension $D$ of the tensors. This approach has already been successfully applied in previous studies of the Shastry-Sutherland model, see e.g. Refs.~\cite{PhysRevB.87.115144,corboz14_shastry}. 
The SE for the ground-state energies of the EPP and FPP were performed by the L\"owdin algorithm~\cite{doi:10.1063/1.1748067, doi:10.1063/1.1724312, PhysRev.77.413} while the energies of the elementary triplon excitations \cite{schmidt03, suppl18} and the dynamic structure factors have been determined using perturbative continuous unitary transformations (pCUTs)~\cite{refId0, 0305-4470-36-29-302}. In all cases we introduce a deformation parameter $\lambda$ so that the unperturbed part $\lambda=0$ corresponds to isolated (empty or filled) plaquettes and $\lambda=1$ to the distorted Shastry-Sutherland model under study.
The ground-state energy for the EPP (FPP) is calculated up to order $9$ ($8$) in $\lambda$. The excitation energies of single triplons have been determined up to order $6$ in both plaquette phases while for the specific one-dimensional case of the orthogonal-dimer chain ($J_2^\prime=0$) order $8$ has been reached. The static structure factors are calculated up to order five. The derived orders are similar to other plaquette expansions~\cite{doi:10.1143/JPSJ.70.1369,PhysRevB.65.014408, PhysRevB.78.224415, PhysRevB.84.134426}. For the distorted Shastry-Sutherland model we increased the maximal perturbative order of the ground-state energies by two compared to Ref.~\onlinecite{Takushima01}.  To our knowledge the dynamic structure factor was not calculated before. All series are extrapolated up to $\lambda=1$ using Pad\'e extrapolation~\cite{guttmann}. In the following we use the variance of the different Pad\'e extrapolants as uncertainty of the extrapolation. For details about both methods, 
see the Supplemental Material~\cite{suppl18}.

%
\begin{figure*}[t]
\begin{center}
\includegraphics[width=0.92\textwidth, trim={3.5cm 23.9cm 3.2cm 2cm}]{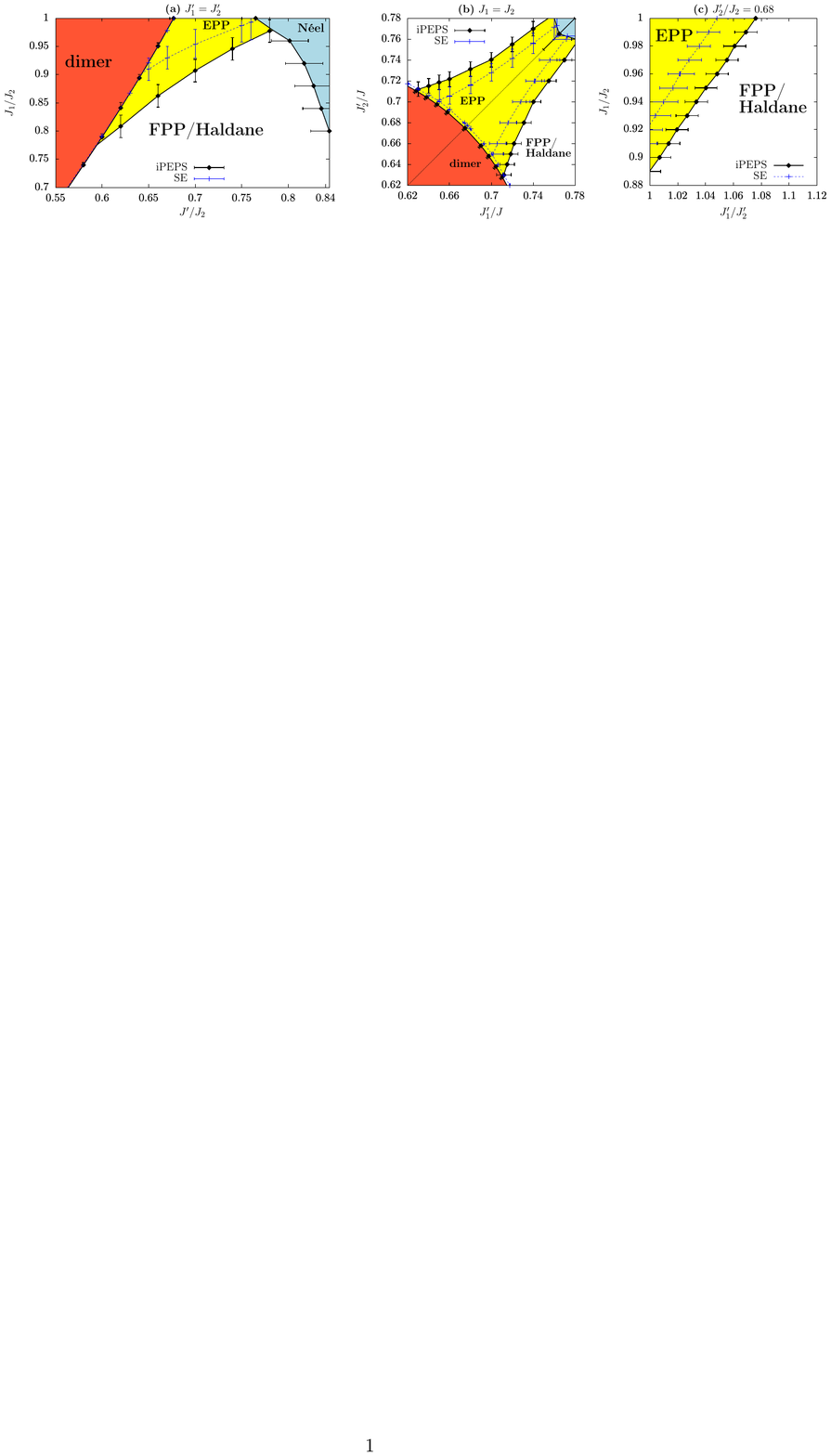}
\caption{Ground-state phase diagrams of distorted Shastry-Sutherland models by iPEPS and SE. In (a) for identical nearest-neighbor couplings $J_1'=J_2'$, in (b) for identical diagonal couplings and in (c) both asymmetries are included at the ratio $J_2'/J_2=0.68$. The area, based on the iPEPS results, of the dimer singlet phase is colored in red, the EPP in yellow, the FPP/Haldane phase in white, and the N\'eel phase is shown in blue.}
\label{Fig2}
\end{center}
\end{figure*}

Since we have four parameters, hence three up to an overall energy scale, plotting the full phase diagram is tricky. We have chosen to study the phase diagram in three planes defined by $J'_1=J'_2=J'$ (Fig.~\ref{Fig2}a), $J_1$=$J_2$=$J$ (Fig.~\ref{Fig2}b), and $J'_2/J_2=0.68$ (Fig.~\ref{Fig2}c). In the phase diagram of Fig.~\ref{Fig2}a, we revisit the effect of different intra-dimer couplings discussed in Ref.~\cite{PhysRevB.83.140414}. Qualitatively, the results are the same, with four phases (dimer, EPP, N\'eel, and Haldane), but the extent of the EPP is considerably reduced, and accordingly the Haldane phase is stabilized in a much larger parameter range that extends up to $J_2/J_1\simeq 0.98$, very close to the isotropic point. In the phase diagram of Fig.~\ref{Fig2}b, we study the effect of different inter-dimer couplings. This phase diagram shows that the EPP is indeed the only one appearing in the isotropic Shastry-Sutherland model, but that it only takes a modest difference to stabilize the FPP. Finally, in Fig.~\ref{Fig2}c, we show a cut in which both the ratios $J_2/J_1$ and $J'_2/J'_1$ vary for a fixed value of $J'_2/J_2=0.68$. From the previous phase diagrams, we know that at the bottom left corner the Haldane phase has to be stabilized, while at the top right corner, the FPP is stabilized. Quite remarkably, there is no phase transition between them, and these two phases actually constitute a single quasi-one dimensional phase in which strong correlations are concentrated around full plaquettes (see Fig.~\ref{Fig1}c). Similar correlations have actually already been reported in the bottom right panel of Fig.~3 of Ref.~\cite{PhysRevB.83.140414}.

To further demonstrate that the FPP and the Haldane phase are adiabatically connected we have computed the inter- and intra-dimer spin-spin correlations and the correlation lengths along a linear path in parameter space connecting the model with unequal inter-dimer couplings ($J_2'/J_2=0.66$, $J_1'/J_2'=1.1$, $J_1/J_2=1$) to the one with unequal intra-dimer couplings  ($J_2'/J_2=0.55$, $J_1'/J_2'=1$, $J_1/J_2=0.5$). The iPEPS results ($D=10$ full update simulations) given in the Supplemental  Material~\cite{suppl18} show that all correlations change smoothly, i.e.~that there is no sign of a quantum phase transition along this path. Interestingly, the ratio of the correlation lengths in x- and y-direction, $\xi_x/\xi_y$, remains almost constant along this path, revealing the anisotropic nature of this phase, that we will now call the FPP/Haldane phase, even in the limit of equal intra-dimer couplings ($J_1=J_2$). 

Let us turn to the properties of the EPP and the FPP/Haldane phase. As mentioned above, the first indication that the EPP phase cannot be the intermediate phase came from NMR~\cite{doi:10.1143/JPSJ.76.073710}, that detected two types of Cu sites. Since NMR is (by necessity) performed in a finite magnetic field, it is interesting to look for complementary evidence in zero-field experiments, ESR, neutron scattering and specific heat~\cite{doi:10.7566/JPSJ.87.033701,Zayed_plaquette_16,sandvik2019}. All these experiments confirm the presence of two well defined magnetic excitations, one at an energy comparable to that of the gap in the dimer phase just before the transition (ESR, neutron scattering), and one at an energy about two times smaller (neutron scattering, specific heat). Neutron scattering could follow the dispersion along the line $(k_x,k_y=0)$ in the Brillouin zone, while specific heat could keep track of the pressure dependence of the gap, i.e. of the minimum of the lowest excitation, with clear evidence that it decreases with pressure. Neutron scattering also revealed that the structure factors of the two low-lying excitations have different momentum dependences.

%
\begin{figure*}[t]
\includegraphics[width=\textwidth, trim={3cm 22.cm 3.2cm 2cm}]{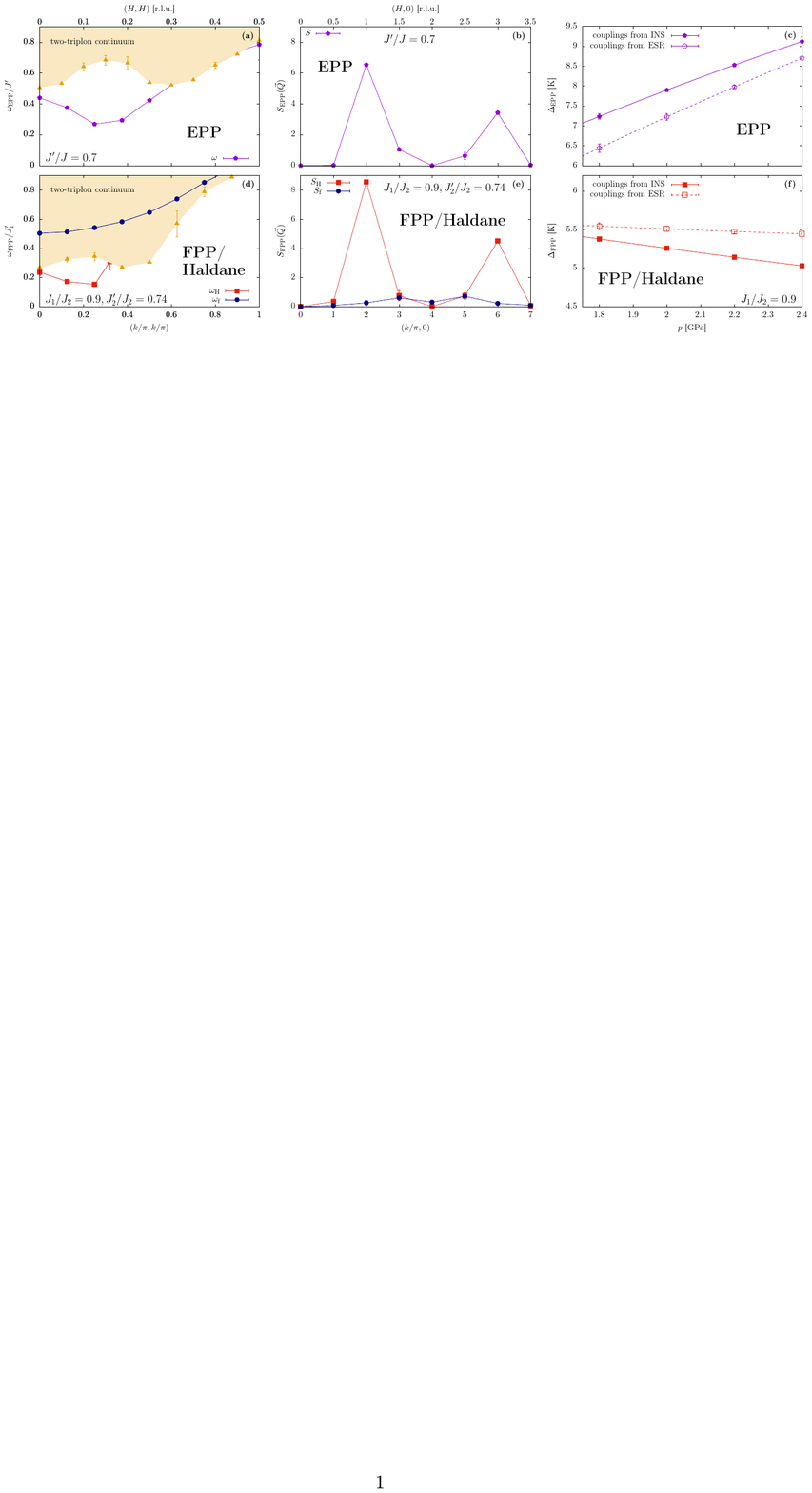}
\caption{Magnetic excitations in the EPP (top) and FPP/Haldane (bottom). Panels (a) and (d): magnetic excitations along $k_x=k_y$; panels (b) and (e): static structure factors along $k_y=0$. The parameters are given inside the figures. Panels (c) and (f): pressure dependence of the gap (with couplings from INS~\cite{Zayed_plaquette_16} and ESR~\cite{doi:10.7566/JPSJ.87.033701}; see main text). All lines are guides to the eye.}
\label{Fig3}
\end{figure*}

To make contact with these experiments, we have studied the magnetic excitations in both phases. Characteristic results are plotted in the top panels (Fig.~\ref{Fig3}a-c) for the EPP and in the bottom panels (Fig.~\ref{Fig3}d-f) for the FPP. In the EPP there is a single low-energy excitation that can be interpreted as the dispersion of a triplet plaquette in a sea of singlet plaquettes. For small to intermediate ratios $J'/J$ its dispersion has a minimum along the direction $k_x=k_y$. Since ESR only measures the zero-momentum excitations while the specific heat detects the gap, i.e.~the minimal energy, this dispersion could be compatible with ESR and specific heat. However, this is not possible if the ratio between both energies is slightly larger than two, since then the mode at $\vec{k}=(0,0)$ decays. The structure factor matches that of the lowest excitation detected in neutron scattering. However, this possibility is excluded by neutron scattering, which has observed two well defined excitations at the same momentum.
Besides, the energy gap in the EPP increases with pressure as shown in Fig.~\ref{Fig3}f, where the pressure is introduced by changing the ratio $J'/J$ in the isotropic model following~\cite{Zayed_plaquette_16, doi:10.7566/JPSJ.87.033701}. This is in clear contradiction with specific heat data.
Note that this remains true also for the EPP in a model with stronger couplings around one set of empty plaquettes, which corresponds to the intrinsic lattice distortion of that phase~\cite{sachdev2019}.

In the FPP/Haldane phase, the situation is very different. There are two well defined excitations (the upper one does not decay along the diagonal $k_x=k_y$~\cite{suppl18}), and, at least not too far from the N\'eel phase, the lowest one has an energy about half that of the other one at $k=0$. The structure factors of these excitations are in good agreement with neutron scattering~\cite{Zayed_plaquette_16}. In addition, the gap decreases with pressure, which matches with specific heat data. This conclusion has been reached following a path in parameter space assuming $J_1=0.9J_2$ and adjusting their average to the estimates from INS for a symmetric model, but we have checked that the sign of the slope remains negative for similar paths. So the case in favor of the FPP appears to be very strong.

The presence of two low-lying excitations in the FPP/Haldane phase can be traced back to the quasi one-dimensional nature of this phase. In the limit of completely decoupled chains ($J'_2=0$), the branch called Haldane corresponds to the triplet excitation branch of a spin-1 chain, realized when all the weak $J_1$ dimers are in a triplet state, while the branch called flat, which is indeed completely flat in that limit, corresponds to a singlet dimer on one of the weak $J_1$ bonds (see Supplemental Material~\cite{suppl18}).

Let us now briefly discuss the implications of the present results for the intermediate phase of SrCu$_2$(BO$_3$)$_2$. Within the minimal model studied in this paper (a purely 2D model with two types of intra- and inter-dimer couplings), there is a single alternative to the EPP of the Shastry-Sutherland model, namely a quasi-1D phase with strong correlations around full plaquettes, and the properties of this phase appear to be consistent with available data. If the system was purely 2D, the stabilization of this phase would induce an orthorhombic distortion since the $C_4$ symmetry is lost. This can be expected to remain true for SrCu$_2$(BO$_3$)$_2$, which is a 3D crystal, if, in all layers, the weak intra-dimer couplings are oriented in the same direction. However, if this direction alternates from one layer to the next, the distortion is not expected any more to be a clear orthorhombic distortion, but to be some local rearrangement inside an essentially unchanged unit cell. The failure so far to detect any clear lattice distortion in the intermediate phase points rather to the second possibility with alternating directions.

There is also an interesting conceptual difference between the two plaquette phases regarding the nature of the phase transition. The EPP is an instability of the Shastry-Sutherland model that spontaneously breaks the symmetry even if all intra- and inter-dimer couplings remain the same. By contrast, the FPP is not an instability of the Shastry-Sutherland model. Like a spin-Peierls transition in spin-1/2 chains, it has to be an instability of the coupled spin-lattice system. So, when applying pressure, if there is a direct transition between the dimer phase and the FPP, it has to take place below the critical ratio at which the transition to the EPP takes place in the Shastry-Sutherland model. Otherwise, there would first be a transition to the EPP. Current estimates of the ratio $J'/J$ at 1.7 GPa from ESR and susceptibility are in the range 0.66-0.665~\cite{Zayed_plaquette_16, doi:10.7566/JPSJ.87.033701, sandvik2019}, indeed below the critical ratio 0.675 of the EPP.

What could be the next step to confirm (or discard) the FPP as the intermediate phase of SrCu$_2$(BO$_3$)$_2$? Of course, a direct identification of the structural distortion would be ideal, but even if the distortion turns out to be too small to be detected, one could hope to detect it indirectly through selection rules. In that respect, measuring the phonons with Raman scattering as a function of pressure could be very helpful. Alternatively, since in our calculations the details of the excitation spectrum change significantly inside the intermediate phases, additional inelastic neutron scattering measurements would be most welcome. Finally, a theoretical investigation of the properties of the intermediate phase in a magnetic field to make contact with NMR and with magnetization measurements is clearly needed. Work is in progress along these lines.


\begin{acknowledgments}
We acknowledge useful discussions with M. Takigawa, H. R{\o}nnow, A. Zheludev, and D. Badrtdinov.
This work was supported by the Swiss National Science Foundation (SNF).
We acknowledge financial support by the German Science Foundation (DFG) through the Engineering of Advanced Materials Cluster of Excellence (EAM) at the Friedrich-Alexander University Erlangen-N\"urnberg (FAU). This project has received funding from the European Research Council (ERC) under the European Union's Horizon 2020  research and innovation programme (grant agreement No 677061). The calculations have been performed using the facilities of the Erlangen Regional Computing Center (RRZE). 

\end{acknowledgments}

\clearpage
\widetext
%
\begin{center}
\textbf{\large Supplemental Material for "Competition between intermediate plaquette phases in SrCu$_2$(BO$_3$)$_2$ under pressure"}

C. Boos, S.P.G.~Crone, I.A.Niesen, P.~Corboz, K.~P.~Schmidt and F.~Mila

\end{center}

\setcounter{equation}{0}
\setcounter{figure}{0}
\setcounter{table}{0}
\setcounter{page}{1}
\makeatletter
\renewcommand{\theequation}{S\arabic{equation}}
\renewcommand{\thefigure}{S\arabic{figure}}

\thispagestyle{empty}

\section{Infinite projected entangled-pair states}
An infinite projected entangled-pair state (iPEPS)~\cite{verstraete_renormalization_2004,nishio2004,jordan_classical_2008} is an efficient tensor network variational ansatz tailored for systematically approximating  ground states of two-dimensional lattice models in the thermodynamic limit. It can be seen as a natural extension of (infinite) matrix product states to two dimensions. 
On a square lattice, an iPEPS consists of a periodically repeated rectangular unit cell made up of five-legged tensors. Each tensor has a single physical leg representing the local Hilbert space of one or several lattice sites, and four auxiliary legs. The auxiliary legs connect to neighboring tensors such that the whole forms a square lattice network of tensors. The accuracy of the ansatz is systematically controlled by the dimension $D$ of the auxiliary legs, called the bond dimension. The iPEPS ansatz used in the present work consists of a $2\times2$ unit cell, with one tensor per dimer (a similar ansatz was used in setup D in Ref.~\onlinecite{corboz_tensor_2013}). In order to increase the efficiency of the simulations we implement a global U(1) symmetry (a subgroup of the SU(2) symmetry of the model) in the iPEPS tensors, see Refs.~\onlinecite{singh_tensor_2011,bauer_implementing_2011} for details.

Given a Hamiltonian $H$, the goal of an iPEPS simulation is to find the optimal tensors that provide the best possible approximation to the ground state of $H$. The simulation starts from either a randomly initialized or a previously converged state, which is then optimized using imaginary-time evolution by applying the operator $e^{-\beta H}$ to the initial state for sufficiently large $\beta$. The evolution operator is split into a product of two-body operators by means of a Trotter-Suzuki decomposition. Application of the two-body operator to two neighboring tensors increases the dimension of the auxiliary bond connecting them, which then needs to be truncated back to the original dimension $D$. This can be done by using either the simple-~\cite{jiang_accurate_2008} or full-update~\cite{corboz_simulation_2010,phien_infinite_2015} method. The former method truncates the updated bond by applying a singular-value decomposition to the tensors connected to the bond, and keeping only the $D$ largest singular values. This approach is computationally inexpensive, but it does not provide the most optimal truncation. In contrast, the full update takes the whole wave function into account when truncating the updated bond, which is optimal, but computationally more demanding. In the present work we have run simple- and full-update simulations up to $D=10$. We have also crosschecked our results for smaller $D$ using variational optimization~\cite{corboz16b}.

Once the optimized tensors have been obtained, expectation values can be calculated by contracting the infinite two-dimensional network formed by the iPEPS, its complex conjugate and the observable of interest. Since two-dimensional tensor networks cannot be exactly contracted in an efficient way, a controlled, approximate contraction method is required.
In this work the contraction is done with the corner-transfer matrix (CTM) method~\cite{nishino_corner_1996,orus_simulation_2009} adapted for a rectangular unit cell~\cite{corboz_stripes_2011,corboz_competing_2014}. The CTM introduces a new parameter, called the boundary dimension $\chi$, which controls the accuracy of the contraction. We always choose $\chi$ to be large enough such that the error induced by $\chi$ is negligible compared to the error due to the finite $D$. 

To determine the location of the phase boundaries between two phases, we first obtain two initial states biased towards both respective phases. The states are generated by evolving randomly initialized iPEPS in imaginary time with the simple-update method using a biased Hamiltonian. Next, using the biased initial states, we perform multiple simulations in the vicinity of the phase transition. Due to hysteresis, a state initialized in one phase will remain within this phase slightly beyond the transition point. The critical coupling is found by determining the point where the linearly interpolated energies of the two phases intersect. All the phase boundaries in Fig.~2 in the main text have been determined by this procedure, using $D=10$ full-update simulations. 

\begin{figure}
\includegraphics[width=0.8\columnwidth]{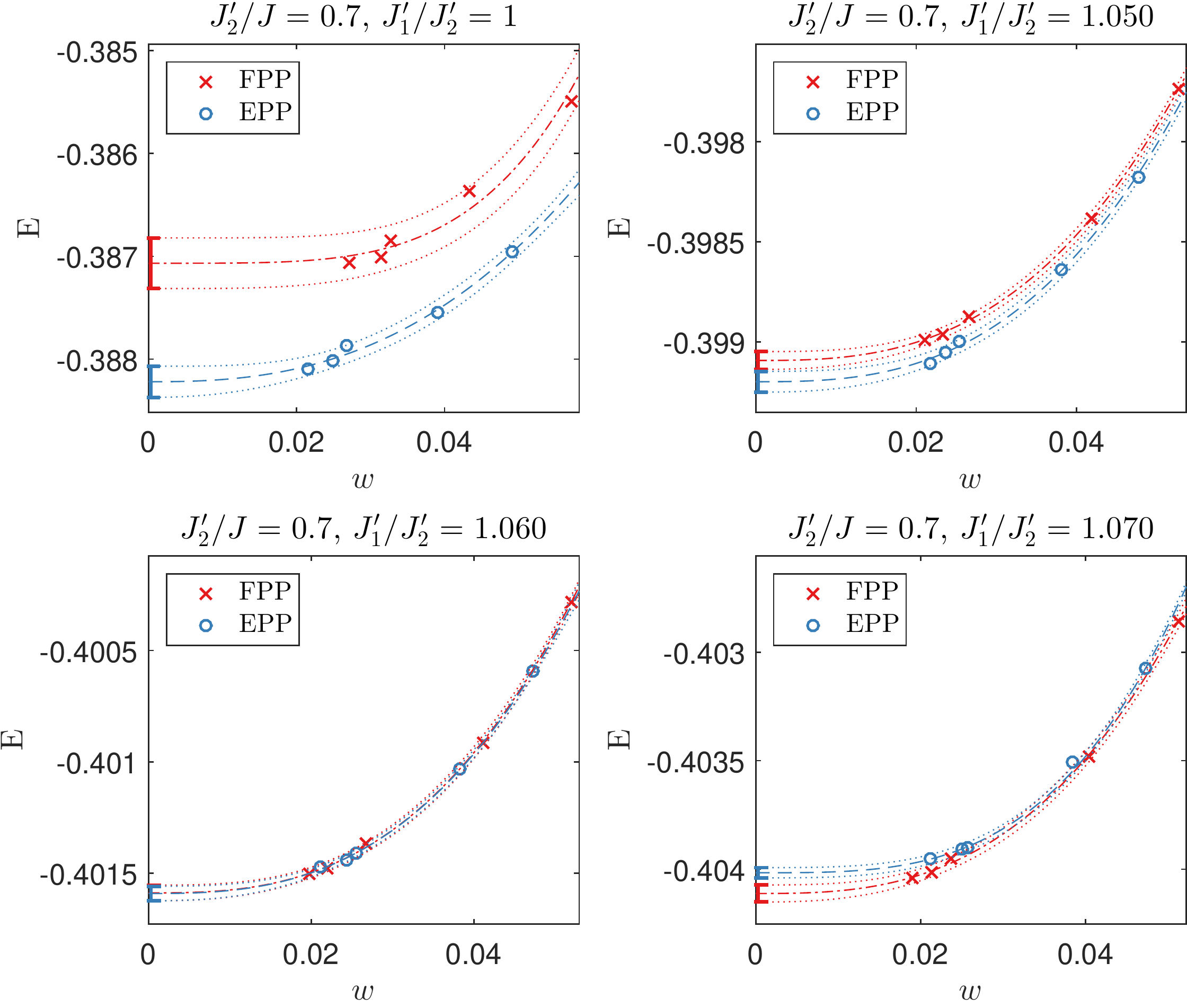}
\caption{Energy per site of the FPP and EPP states as a function of the truncation error $w$ obtained with iPEPS (using the full-update with bond dimensions ranging from $D=6 \ldots 10$), for different values of  $J'_1/J'_2$ and fixed value of $J_2'/J_2=0.7$. The extrapolations are obtained by a polynomial fit of the data, where the shown error is given by the $1\sigma$ confidence interval.  
\label{Fig:iPEPS}}
\end{figure}

We have verified that finite $D$ effects on the phase boundary are small by comparing the finite $D=10$ results to the ones obtained by extrapolating the energies to the exact, infinite $D$ limit, for the transition with fixed $J'_2/J_2=0.7$. The extrapolation is done based on the truncation error $w$ which quantifies the degree of approximation in a state, and goes to zero in the exact limit (see Ref.~\onlinecite{corboz_improved_2016} for details). Figure~\ref{Fig:iPEPS} shows the energies of the two plaquette states for different values of $J'_1/J'_2$. For the unbiased Hamiltonian ($J'_1/J'_2=1$), we find that the empty plaquette phase (EPP) is clearly lower in energy than the filled plaquette phase (FPP). When increasing $J'_1/J'_2$, taking the extrapolated energies including their error bars into account, we find a transition value of $J'_1/J'_2=1.060(8)$. In comparison, the $D=10$ result for the critical coupling is $J'_1/J'_2=1.058$, which is very close to the extrapolated result and lies well within the extrapolation error bar. Thus, we conclude that the $D=10$ result already provides a good accuracy of the phase boundary in the infinite $D$ limit. We take the extrapolation error computed here as a representative error estimate of the phase boundary in Figs.~2(b-c) of the main text.  A similar approach was used to obtain an estimate of the error bar on the phase boundary between the EPP and FPP in vertical direction and between the FPP and N\'{e}el phase in horizontal direction in Fig.~2a. 

The $D=10$ phase boundary between the FPP and dimer phase in Fig.~2a represents a lower bound for the phase transition, since the energy of the dimer phase is known exactly, whereas the energy of the FP state decreases with increasing $D$. The error bar is obtained by intersecting the  energy of the dimer state with the one of the FP state extrapolated in $1/D$. The latter provides a lower bound of the exact ground state energy since the energy converges faster than linearly in $1/D$.  The error estimate between the EPP and dimer phase in Fig.~2a is obtained with a similar approach. The other error estimates in Fig.~2a-b between either the dimer or N\'{e}el phase and the EPP have been obtained from Ref.~\onlinecite{corboz_tensor_2013}.

\subsection{iPEPS results along a path in parameter space}
\begin{figure}[h]
\begin{center}
\includegraphics[width=0.8\columnwidth]{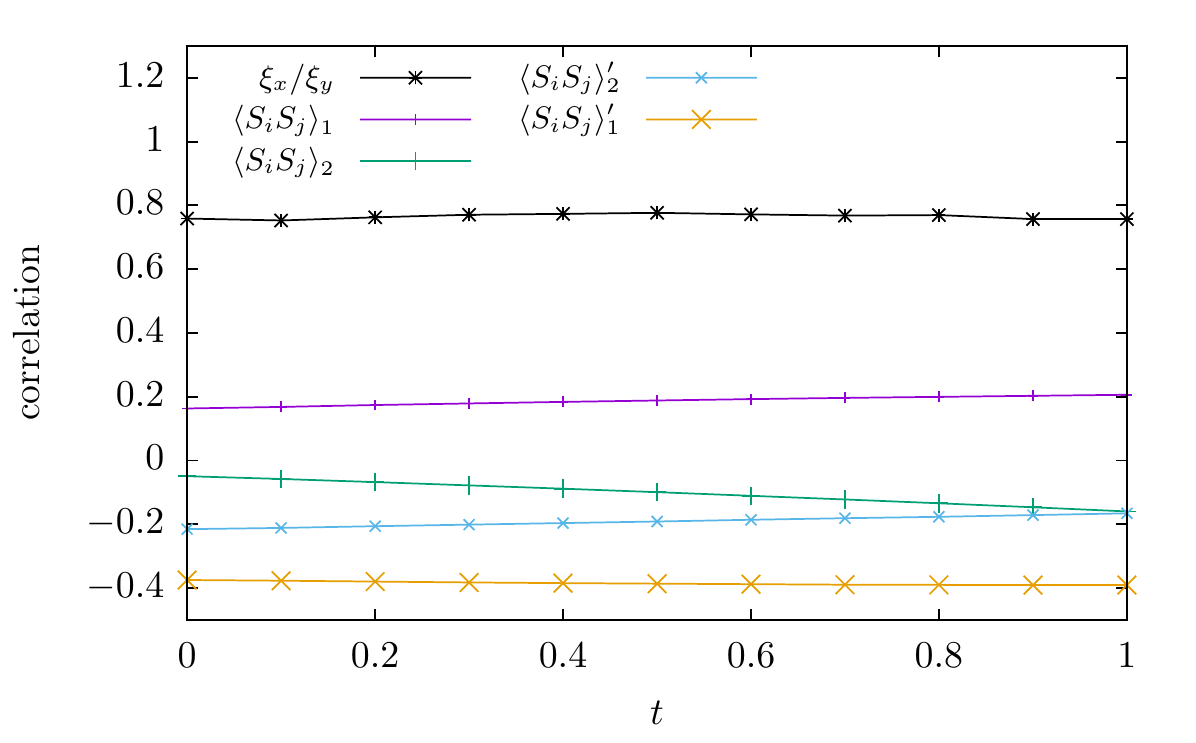}
\caption{Various correlations along a path connecting the model with only asymmetric intra-dimer couplings to the model with only asymmetric inter-dimer couplings. The path is parametrized by $(J_2'/J_2, J_1'/J_2', J_1/J_2) = (0.66 - 0.1t, 1.10- 0.1t, 1 - 0.5t)$ with $t \in [0,1]$. The correlations with a prime $\langle \bullet \rangle '$ refer to inter-dimer bonds, the ones without a prime $\langle \bullet \rangle$ to the intra-dimer bonds. Additionally, the anisotropy in the correlation length $\xi_x/\xi_y$ is shown. All correlations change smoothly, i.e. there is no sign of a quantum phase transition along this path, from which we conclude that the FPP and Haldane phase are the same phase which we call the FPP/Haldane phase.
}
\label{Fig:correlations}
\end{center}
\end{figure}

In Fig.~\ref{Fig:correlations} we present the iPEPS results for the inter- and intra-dimer spin-spin correlations and the correlation length along the path in parameter space mentioned in the main text, i.e. a linear path connecting the model with unequal inter-dimer couplings ($J_2'/J_2=0.66$, $J_1'/J_2'=1.1$, $J_1/J_2=1$) to the one with unequal intra-dimer couplings  ($J_2'/J_2=0.55$, $J_1'/J_2'=1$, $J_1/J_2=0.5$).  These results have  been obtained by $D=10$ full-update simulations, starting from a state in the FPP. 
 To exclude that the results are biased due to this choice of the initial state we have verified that the results at the end point of the path in the Haldane phase agree with the ones obtained when starting from randomly initialized tensors. We further note that the correlation lengths have been computed based on the largest and second largest eigenvalue of the row-to-row transfer matrix along both directions, as explained in Ref.~\cite{nishino96}.

%

\section{Series expansion methods}
\subsection{Ground-state energies}
%
%
\begin{figure}[b!]
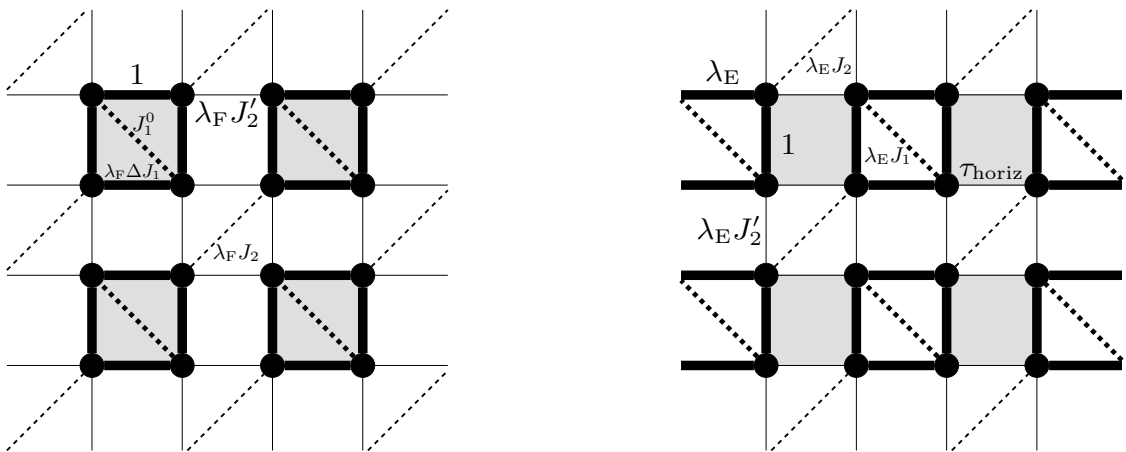

\begin{minipage}{0.5\columnwidth}
  \centering
  \scalebox{0.6}{\shastrysutherlandlatticefullapproachpCUTEzero{}}\\
\end{minipage}%
\begin{minipage}{0.5\columnwidth}
  \centering
  \scalebox{0.6}{\shastrysutherlandlatticeemptyapproachpCUT{}}\\
\end{minipage}
\caption{Illustration of the models for the SE approaches. In the left panel the deformed Hamiltonian for the expansion about the full plaquette singlet state is indicated.
The deformed Hamiltonian for the expansion of the empty plaquette state is shown in the right panel.
The unperturbed plaquettes at $\lambda_{\text{F}} = 0$ or $\lambda_{\text{E}} = 0$ are shaded in gray.
The exchange $\tau_{\text{horiz}}=1+\lambda_{\text{E}}(J_2'-J_1')$ introduces the asymmetry on the empty plaquettes hosting singlets.}
\label{Fig:ex_shastry_model_approaches_E0}
\end{figure}
A subtle point about plaquette phases in the distorted Shastry-Sutherland model is given by the spatial location of the dressed plaquette singlets. In the following we simply refer to them as singlets.
At first there is the obvious distinction between the location on either plaquettes containing or not containing an inner diagonal dimer coupling, corresponding to the FPP or to the EPP, respectively. Besides, in each case, there is the choice of the subset of plaquettes on which the singlets are formed.
In the case of the FPP, the singlets can either be on plaquettes formed by only $J_1'$ or $J_2'$ couplings.
For identical dimer couplings $J_1=J_2$ with the bias $J_1'>J_2'$ the singlets are located on the $J_1$ plaquettes, whereas for $J_2'>J_1'$ the $J_2$ plaquettes host the singlets.
Therefore, the FPP/Haldane phases shown in the Fig.~2b of the main article actually represent two distinct phases which are separated by intermediate phases in this parameter range. For other parameters with $J_1 \neq J_2$ this is not necessarily the case, and in particular for $J_2<J_1$ and $J_1'>J_2'$ or vice versa, a direct phase transition between these two FPP/Haldane phases occurs. By contrast, the possibility to choose between two sets of empty plaquettes leads to a single EPP phase in  Fig.~2a-c of the main article with a two-fold degenerate ground state.

For the ground-state energies per site $\epsilon_0$ we apply perturbation theory after L\"owdin~\cite{doi:10.1063/1.1748067, doi:10.1063/1.1724312, PhysRev.77.413} and perform linked-cluster expansions (LCEs).
%
The series expansion (SE) about the FPP is performed from an unperturbed model of decoupled filled plaquettes at $\lambda=0$. In the following we set the inter-dimer coupling of these plaquettes $J_1'=1$ and choose an intra-dimer coupling $J_1$ of intermediate strength $J_1^0=1/0.74$ at $\lambda=0$. The intra-dimer coupling in the physical model at $\lambda = 1$ is then reached by an additional local perturbation on the intra-dimer bond. This approach for the FPP with such an initial value of $J_1^0$ turns out to be efficient to study a broad range of values, and is especially well converged in the parameter regime of interest. The Hamiltonian used for the SE of the FPP reads
\begin{align}
\mathcal{H}^{\text{FPP}}&= \sum_{\substack{\langle i,j \rangle \\ \text{bold}}} \vec{S}_i  \cdot \vec{S}_j
+ J_1^0 \sum_{\substack{\langle\langle i,j \rangle\rangle \\ \text{bold}}} \vec{S}_i \cdot \vec{S}_j
+ \lambda_{\text{F}} J_2' \sum_{\substack{\langle i,j \rangle \\ \text{thin}}} \vec{S}_i  \cdot \vec{S}_j
+ \lambda_{\text{F}} \Delta J_1 \sum_{\substack{\langle\langle i,j \rangle\rangle \\ \text{bold}}} \vec{S}_i \cdot \vec{S}_j
+ \lambda_{\text{F}} J_2 \sum_{\substack{\langle\langle i,j \rangle\rangle \\ \text{thin}}} \vec{S}_i \cdot \vec{S}_j,
\label{eq:ham_E0_expansion_FPP}
\end{align}
where the sums can be understood with the lattice given in the left panel of Fig.~\ref{Fig:ex_shastry_model_approaches_E0}. The diagonal couplings of the initial plaquettes are tuned to the physical value by $\Delta J_1 = J_1 -J_1^0$.

For the EPP the unperturbed model at $\lambda = 0$ is given by symmetric empty plaquettes. The inter-plaquette interactions are introduced perturbatively. Locally, the empty plaquettes in the distorted Shastry-Sutherland model exhibit different strengths on neighboring bonds of the plaquette, which is why an additional local interaction is introduced on two opposing bonds on the empty singlet plaquettes. The Hamiltonian used for the SE of the EPP is
\begin{equation}
\begin{aligned}
\mathcal{H}^{\text{EPP}}&=
 \sum_{\substack{\langle i,j \rangle \\ \text{vertical} \\ \text{bold}}} \vec{S}_i \cdot \vec{S}_j
 + \left(1+\lambda_{\text{E}}(J_2'-1)\right) \sum_{\substack{\langle i,j \rangle \\ \text{horizontal} \\ \text{thin}}} \vec{S}_i \cdot \vec{S}_j
+ \lambda_{\text{E}} \sum_{\substack{\langle i,j \rangle \\ \text{horizontal} \\ \text{bold}}} \vec{S}_i \cdot \vec{S}_j
+ \lambda_{\text{E}} J_2' \sum_{\substack{\langle i,j \rangle \\ \text{vertical} \\ \text{thin}}} \vec{S}_i \cdot \vec{S}_j\\
& + \lambda_{\text{E}} J_1 \sum_{\substack{\langle \langle i,j \rangle \rangle \\ \text{thin}}} \vec{S}_i \cdot \vec{S}_j
 + \lambda_{\text{E}} J_2 \sum_{\substack{\langle \langle i,j \rangle \rangle \\ \text{bold}}} \vec{S}_i \cdot \vec{S}_j.
\end{aligned}
\label{eq:ham_E0_expansion_EPP}
\end{equation}
Note, that the parameter spaces of the two deformed Hamiltonians linking the unperturbed product states of filled and empty plaquette singlets with the adiabatically connected states in the distorted Shastry-Sutherland model are not the same if $\lambda \neq 1$.
We therefore refer to the expansion parameter in the expansion of the FPP with $\lambda_{\text{F}}$ and to the expansion parameter in the expansion of the EPP with $\lambda_{\text{E}}$.

The perturbative description effectively takes place on a lattice of supersites, which are either given by one set of empty or of filled plaquettes.
For all exchanges between these supersites transitions within a basis of 256 nearest-neighbor two-plaquette states occur.
We exploit the linked-cluster theorem and perform a full graph expansion, where bonds of the same coupling need to be distinguished for different directions. For instance a trimer of three supersites has differing contributions for the direction $(2,0)^T$ and $(1,1)^T$ and hence needs to be calculated several times.
This increases the number of required graphs for the ground-state energy $\epsilon_0$ and one has to calculate the energy on 2849 graphs for the expansion in the FPP up to order eight in $\lambda_{\text{F}}$ on a directed triangular lattice and 212 graphs on a directed square lattice for the expansion in the EPP in $\lambda_{\text{E}}$ also up to order eight.
%
%
As a consequence the expansions cannot be pushed to similarly high orders as known for other models with smaller local Hilbert spaces of the supersites~\cite{PhysRevB.65.014408, PhysRevLett.105.267204, PhysRevB.94.201111}. For both special cases $J_1=J_2$ and $J_1'=J_2'$ order nine for the ground-state energies $\epsilon_0^\text{(E)}$ of the EPP is calculated. In that case, separate calculations on 244 graphs are required.

\begin{figure}[b!]
\begin{center}
\includegraphics[width=0.85\columnwidth]{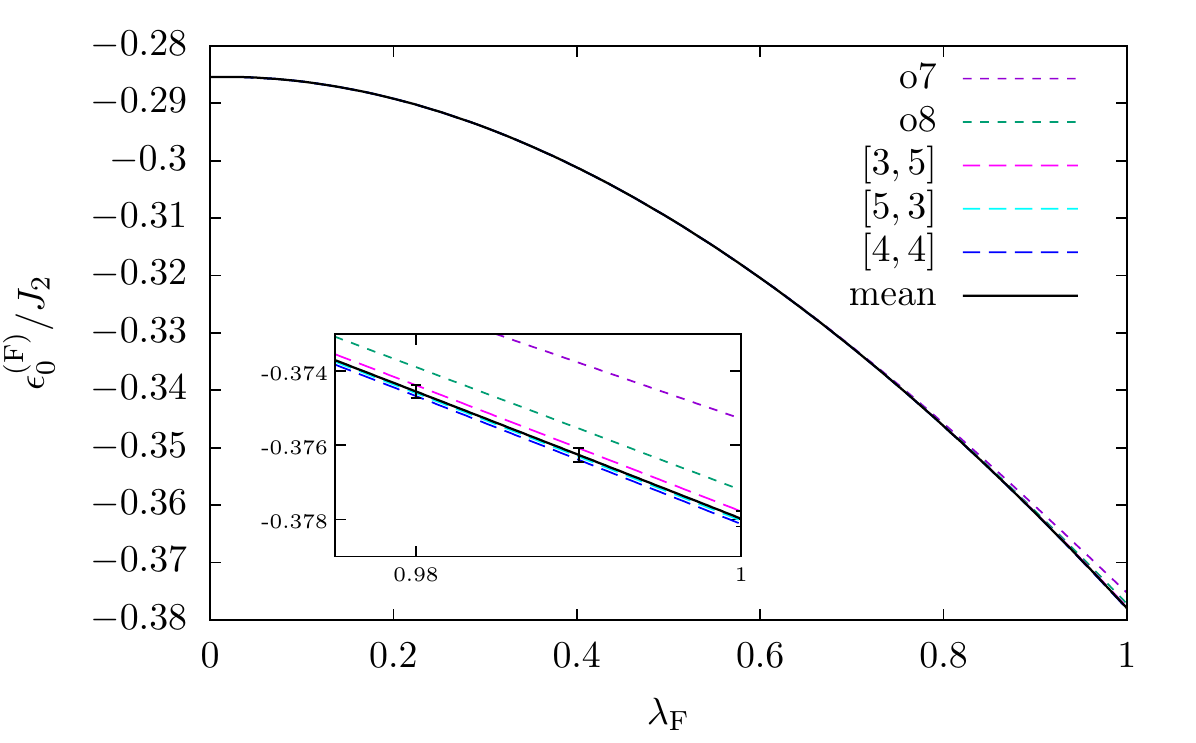}
\caption{Ground-state energies as bare series, Pad\'e extrapolants and mean values of the expansion of the FPP at \mbox{$J_2'/J_2=0.68$}, $J_1'/J_2' = 1.01$ and $J_1/J_2=0.98$.
The point $\lambda_{\text{F}}=1$ belongs to the physically relevant distorted Shastry-Sutherland model.}
\label{Fig:convergence_E0FPP}
\end{center}
\end{figure}
\begin{figure}[t!]
\begin{center}
\includegraphics[width=0.85\columnwidth]{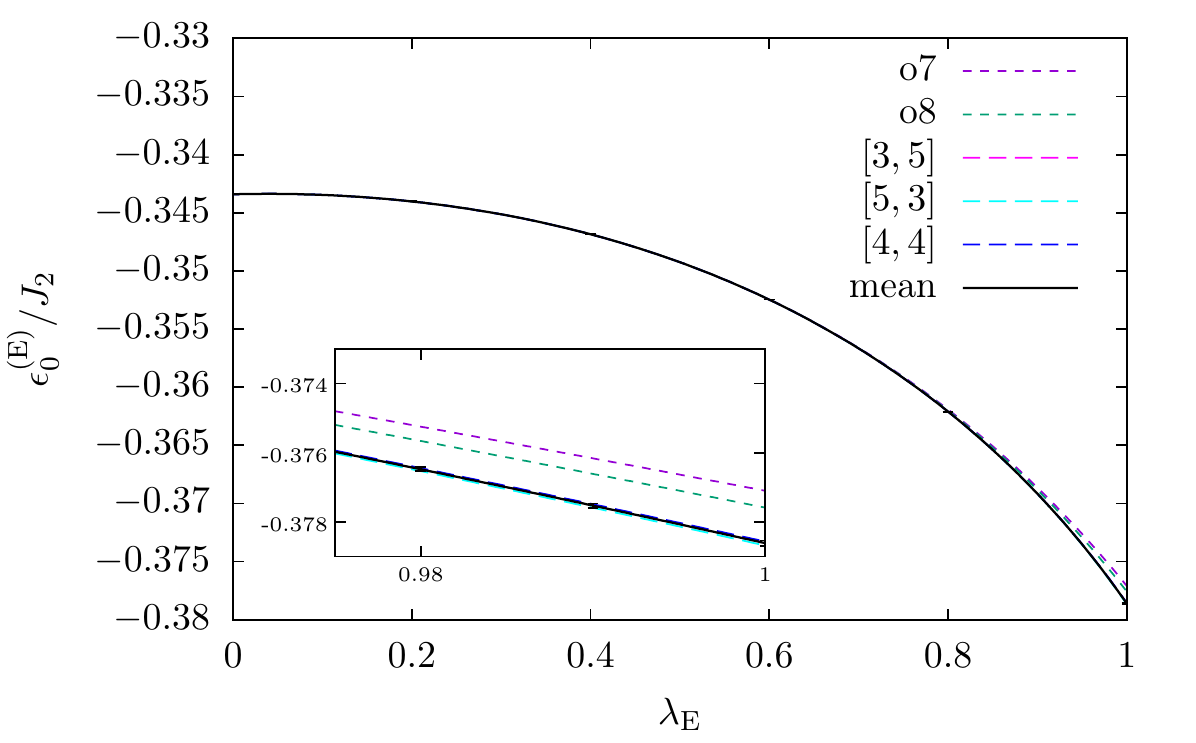}
\caption{Ground-state energies as bare series, Pad\'e extrapolants and mean values of the expansion of the EPP at $J_2'/J_2=0.68, J_1'/J_2' = 1.01$ and $J_1/J_2=0.98$.
The point $\lambda_{\text{E}}=1$ belongs to the physically relevant distorted Shastry-Sutherland model.}
\label{Fig:convergence_E0EPP}
\end{center}
\end{figure}

The polynomial series derived for the ground-state energies of both plaquette phases need to be analyzed carefully with respect to their convergence behavior.
We apply Pad\'e extrapolations which are given by rational functions. They are defined such that the Taylor expansion of the extrapolation equals the original series in the prevailing order~\cite{guttmann}. The exponents of the numerator and denominator polynomials are referred to as $[l,m]$.
An important issue for the usage of these extrapolations are spurious poles. If such a pole arises for an extrapolation in the parameter space of interest or in the close vicinity, the extrapolation has to be excluded from the physical analysis.
The convergence behavior of the Pad\'e extrapolations is analyzed by grouping them into families. All members of a family are characterized by the same difference $l-m$.
If the extrapolations within one family show a convergent behavior, the member with the highest available order is taken as the best converged representant. The average about these representants from different families is considered as the most reliable result. In the following, we use the standard deviation of these extrapolants as a measure for the uncertainty. It is usually plotted as error bar.
Another advantage of the Pad\'e extrapolations as compared to the bare series is that they are better suited to control divergences for large values of $\lambda$. This is in particular true for all extrapolations with similar exponents in the numerator and denominator and the diagonal extrapolations with $l=m$ are expected to yield the most accurate results.
%
%
%
In the following, we analyze the convergence behavior for the ground-state energies of the FPP and EPP and state the detailed choices of Pad\'e extrapolants.
The main guideline is to average over extrapolations in the highest available order of every convergent family with similar exponents $l$ and $m$.

For the pCUT expansion the dependencies of the FPP and EPP ground state energies on the parameters $\lambda_{\text{F}}$ and $\lambda_{\text{E}}$ at the coupling strengths $J_2'/J_2=0.68$, \mbox{$J_1'/J_2' = 1.01$} and $J_1/J_2=0.98$ are shown in Fig.~\ref{Fig:convergence_E0FPP} and Fig.~\ref{Fig:convergence_E0EPP}, respectively.
Displayed are the bare series in orders seven and eight, the Pad\'e extrapolants [3,5], [5,3] and [4,4] as well as the resultant mean value with the corresponding standard deviation taken from this set of extrapolants.
%
\begin{figure}[h!]
\begin{center}
\includegraphics[width=0.8\columnwidth]{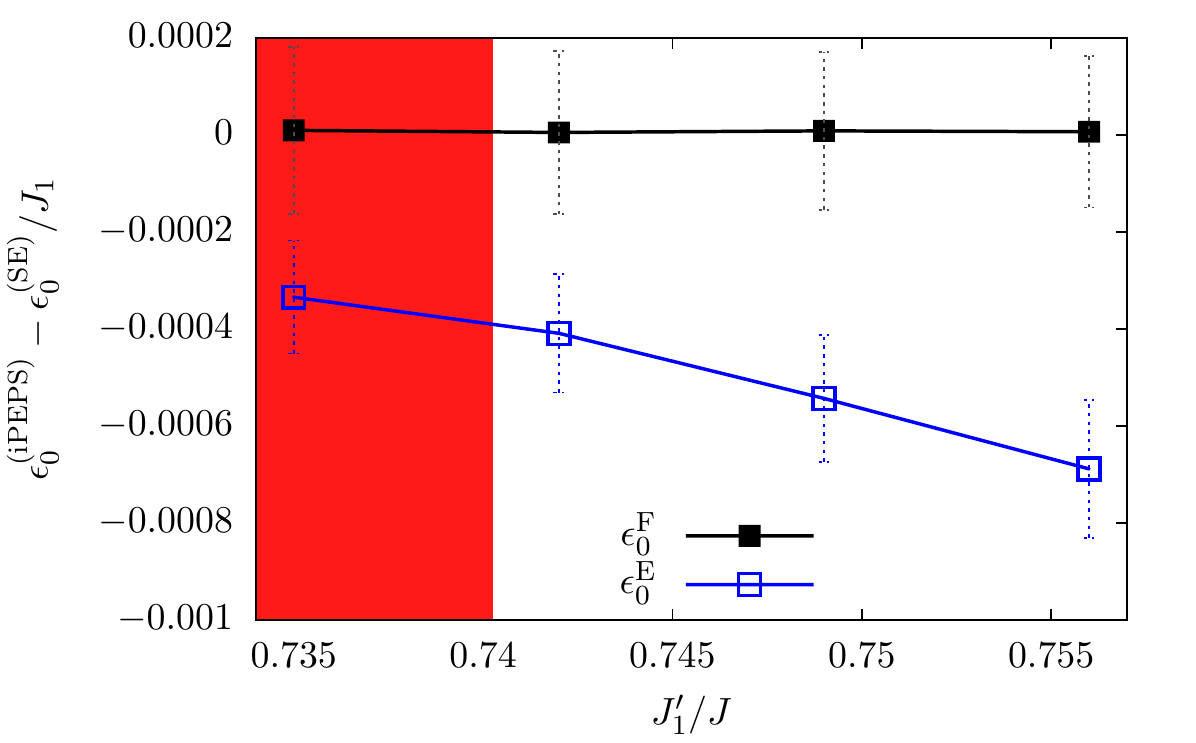}
\caption{Energy differences between ground-state energies from iPEPS and SE along the line $J_2'/J_2=0.7$ with $J_1/J_2 = 1$. The red background color indicates where the EPP is the ground state and no background color signals the FPP as found by iPEPS.}
\label{Fig:convergence_E0cut}
\end{center}
\end{figure}
%
The energies of both states decrease with increasing expansion parameters due to quantum fluctuations. The difference between the unperturbed energy and the energy in the distorted Shastry-Sutherland model is more than twice as large for the FPP than for the EPP. This is due to the fact that the intra-dimer bond on a single plaquette increases the ground-state energy.
Furthermore, the energy splitting between the bare series as well as the Pad\'e extrapolants is larger for the FPP than for the EPP at these coupling values at $\lambda=1$. The standard deviation of both phases becomes more similar in the parameter space where the FPP is the ground state. This can be seen in Fig.~\ref{Fig:convergence_E0cut}, where the energy differences between the SE and the iPEPS results are shown along the line $J_2'/J_2=0.7$ in the case $J_1/J_2 = 1$.
For the EPP in this more symmetric case the series are reached up to order nine in $\lambda_{\text{E}}$ and for the mean value the extrapolants [4,4], [4,5] and [5,4] are used.
The FPP energies of both approaches agree extremely well, whereas for the EPP the difference between iPEPS and SE energies is larger with a value of $\approx 0.0005$.
This difference increases with the ratio $J_1'/J_1$ because the asymmetry on the EPP supersites becomes stronger, which corresponds to a larger perturbation.
The point in parameter space at $\lambda=0$ is noticeably closer to the distorted Shastry-Sutherland model of interest for the FPP than for the EPP. This is due to the fact that one of the strong intra-dimer couplings is mainly included in the unperturbed part of the FPP expansion.
In this sense the perturbation for the EPP is larger, which might lead to an error on the EPP energies that is not reflected in the standard deviation of the Pad\'e extrapolants, but would rather require higher-order calculations to become evident.

For the analysis of the phase diagram in Fig.~2a in the main text we use the mean values of the energies as defined above in order eight for the FPP and order nine for the EPP. The same mean value for the FPP was used to extract the phase diagram in Fig.~2b. For the EPP energies the Pad\'e extrapolant with the exponents [5,4] has to be excluded. In the phase diagram of the completely distorted Shastry-Sutherland model given in Fig.~2c in the main text both energies are taken as mean values of Pad\'e extrapolants in order eight. The parameters where the standard deviations of the EPP and FPP expansion overlap are indicated as error bars.

\subsection{Magnetic excitations}
The magnetic excitations are calculated with SE using perturbative continuous unitary transformations (pCUTs)~\cite{refId0, 0305-4470-36-29-302}, which allows to apply a LCE as well. For the pCUT the unperturbed problem is required to have an equidistant energy spectrum bounded from below. Considering an isolated plaquette, these conditions are fulfilled for an intra-dimer $J_1$-bond with $J_1^{0}=0$ or $J_1^{0}=1$.
It is then possible to rewrite the unperturbed part $\mathcal{H}_0$ of isolated plaquettes in Eq.~\eqref{eq:ham_E0_expansion_FPP} for both cases as a counting operator $\mathcal{Q}$ up to an additional constant $E_0$.
The perturbation $\lambda\mathcal{V}$ can be decomposed as a sum of operator $T_n$ with $n\in\{-4,\ldots,+4\}$ with $[T_n,\mathcal{Q}]=nT_n$, i.e.~the operators $T_n$ change the number of energy quanta by $n$. These energy quanta are called quasi-particles (QPs).
The deformed Hamiltonians can be written as
\begin{equation}
 \mathcal{H} = E_0+ \mathcal{Q} + \lambda\sum_{n=-4}^{4} T_n \quad .
\end{equation}
Within pCUTs, this type of Hamiltonian is unitarily mapped, order by order in $\lambda$, to an effective Hamiltonian $\mathcal{H}_{\rm eff}$ which conserves the number of QPs, i.e.~$[\mathcal{H}_{\rm eff},\mathcal{Q}]=0$. We note that this step can be done model-independently. The model-dependent part of pCUTs corresponds to a normal-ordering of the effective Hamiltonian in the QP sector of interest. This is done most efficiently via a full-graph decomposition using the linked-cluster theorem~\cite{Coester2015}. 

Here we focus on the one-QP sector with total spin one, which is the relevant sector for the comparison with inelastic neutron scattering (INS) measurements.
Such magnetic three-fold degenerate triplet excitations above a magnetically disordered singlet ground state, which are adiabatically connected to singlet-triplet excitations on isolated plaquettes, are called triplons in this work. This therefore generalizes the original terminology, where triplons are dressed triplet excitations adiabatically connected to singlet-triplet dimer excitations~\cite{schmidt03}. 

For the FPP we use $J_1^{0}=1$, where two one-QP triplon modes are present.
In practice, one determines all hopping amplitudes for these excitations up to some order in $\lambda_{\rm F}$ by calculating matrix elements of $\mathcal{H}_{\rm eff}$ between one-triplon states.
The resulting one-QP hopping Hamiltonian can be further digonalized by a Fourier transformation exploiting the translational symmetry of the lattice. 
Starting from the unperturbed plaquette with $J_1^{0}=1$, one gets the $2\times 2$ matrix $\Omega(\vec{k})$ for each $\vec{k}$, which is easily diagonalized yielding two one-triplon dispersions $\omega_{\pm}(\vec{k})$.
These one-triplon dispersions are determined up to order six in $\lambda_{\rm F}$ for the distorted Shastry-Sutherland model.
At specific momenta the two one-triplon modes are protected by the local symmetry on the $J_1$-bonds so that $\Omega(\vec{k})$ is directly diagonal.
These decoupled modes are referred to as $\omega_{\text{H}}(\vec{k})$ and $\omega_{\text{f}}(\vec{k})$.
In addition, we check whether the one-triplon modes have an infinite life-time by calculating the lower band edges $\omega_{\text{lowerbandedge}}^{ab}$ of the two-triplon continua of triplons $a, b\in \{\pm\}$ . The lower band edge can be determined by
\begin{align}
\label{eq_2qp}
\omega_{\text{lowerbandedge}}^{ab}(\vec{k}) = \text{min}_{\vec{q}} \left[ \omega^{a}\left(\frac{\vec{k}}{2}+\vec{q}\right) + \omega^{b}\left(\frac{\vec{k}}{2}-\vec{q}\right) \right].
\end{align}

For the EPP we use $J_1^{0}=0$, where a single one-QP triplon is present in the perturbation theory. In this case the Fourier-transformation directly leads to the dispersion $\omega(\vec{k})$. There is a single two-triplon continuum following Eq.~\eqref{eq_2qp} with $\omega^{a}=\omega^{b}=\omega$.
If a one-triplon mode does not decay for a given wave vector $\vec{k}$, the calculated series can again be extrapolated by Pad\'e approximation~\cite{guttmann}.


For the magnetic excitations the convergence behavior is generically worse than for the ground-state energies.
This has several reasons. Firstly, we only reach the series up to order six.
Secondly, both triplon modes mix typically and therefore more quantum fluctuations contribute. For the specific momenta with $k_x=k_y$ as well as $(k_x=\pm\pi,k_y=0)$ and $(k_x=0,k_y=\pm\pi)$ the two triplon modes are protected, which is why we focus on these values.
In Fig.~3a in the main text for the FPP the Pad\'e extrapolents [2,3] and [3,2] are used. The Pad\'e extrapolation with exponents [3,3] shows unphysical divergences in the chosen parameter space. For the EPP combinations of the extrapolants with the exponents [2,3], [3,2] and [3,3] are used depending on the convergence behavior at the specific momentum.


\subsection{Magnetic dynamic structure factor}
The magnetic dynamic structure factor $S(\vec{k},\omega)$ gives the intensities measured in INS in the approximation of linear response theory.
It reads
\begin{equation}\label{eq::dsf}
\begin{aligned}
S(\vec{k},\omega)&=-\frac{1}{\pi} \Im \left( \left\langle 0 \Big| \mathcal{O}^\dagger(\vec{k}) \frac{1}{\omega-(H-E_0)} \mathcal{O}(\vec{k}) \Big| 0 \right\rangle \right),\\
\mathcal{O}(\vec{k})&= \sum\limits_i \sum\limits_\alpha S^\alpha(\vec{x}_i) \text{e}^{i \vec{k} \vec{x}_i},
\end{aligned}
\end{equation}
where the ground state is denoted with $\ket{0}$.
%
\begin{figure}[b!]
\begin{center}
\includegraphics[width=1.\columnwidth, trim={3cm 21.7cm 3cm 2cm},clip]{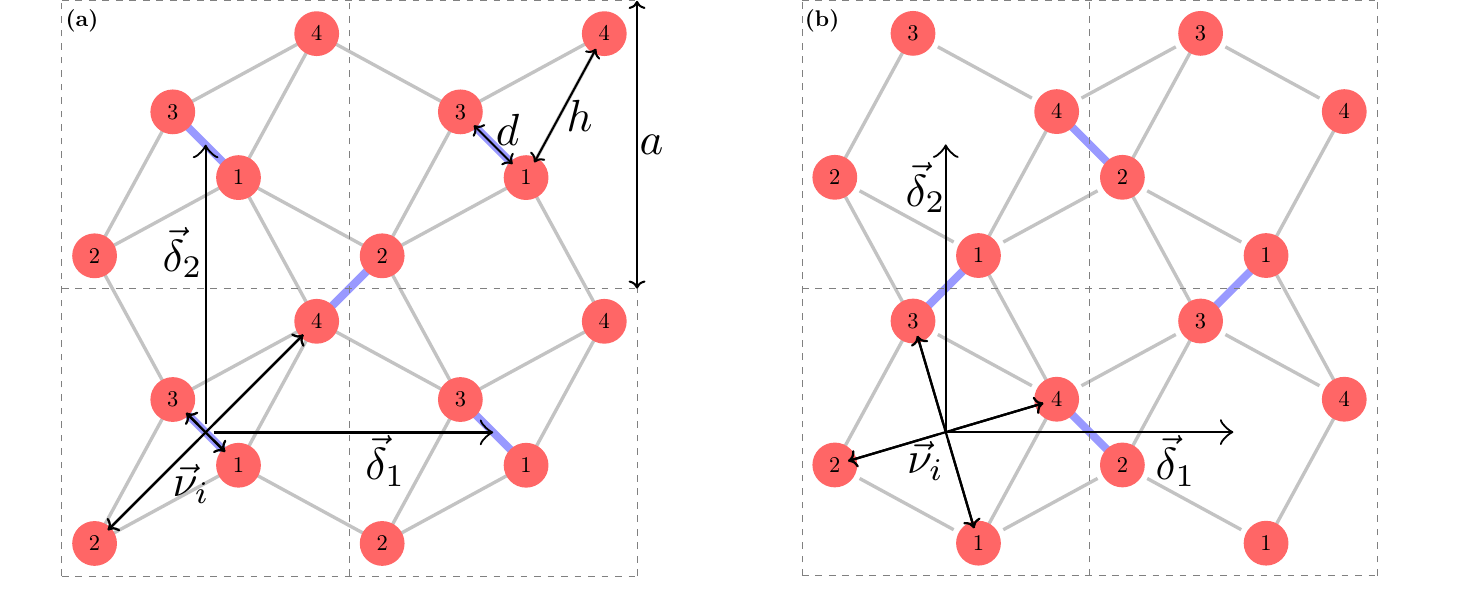}
\caption{Illustrations of the Shastry-Sutherland lattice.
The distances $a$, $d$, and $h$ are given in the text.
The spins are labeled by the position within their plaquette $\nu\in\{ 1,2,3,4 \}$.
The vectors from the plaquette center to the spins are plotted and indicated as $\vec{\nu}_i$.
The lattice vectors are also given as $\vec{\delta}_1$ and $\vec{\delta}_2$.
On the left the unit cell is chosen to be centered around a filled plaquette with a diagonal coupling between the top left spin and the bottom right spin.
On the right one of the choices for the empty plaquettes is shown.
}
\label{Fig:latticeanduc}
\end{center}
\end{figure}
The operator $\mathcal{O}(\vec{k})$ is the Fourier-transformation of the sum of spin operators $S^\alpha$, with $\alpha \in \{x,y,z\}$.
Since the problem is SU(2) invariant it is sufficient to study only $\alpha=z$ and multiply the result by three.
In our approach we consider the problem in terms of 4-site plaquettes and it is most convenient for the Fourier-transformation to label the position of a spin at $\vec{x}_i$ by the plaquette $p$ it belongs to and the position within the plaquette $\nu\in\{1,2,3,4\}$.
For the comparison with experiments the real lattice structure has to be taken into account.
It is illustrated in Fig.~\ref{Fig:latticeanduc} including the real space distances, which have been measured to be $a=8.99\angstrom$, $h=5.120\angstrom$, and $d=2.905\angstrom$~\cite{SMITH1991430}. For convenience we define $\tilde h = \sqrt{\left(h^2-\left(\frac{d}{2}\right)^2 \right) / 2}$.
The crystal vectors are
\begin{equation}
\begin{aligned}
\vec{\delta}_1&=a(1,0)^T, \qquad  \vec{\delta}_2=a(0,1)^T.
\end{aligned}
\end{equation}
The lattice offers two distinct orientational choices for both the empty and the filled plaquettes.
Here we choose a single one, under the expectation that the symmetry is broken by the realization of either of the plaquette singlet phases, leading to one distinct choice.
The other orientation leads to the same structure factors only rotated in momentum space ($k_x\rightarrow k_x, k_y\rightarrow -k_y$).
The inclusion of both orientations is only necessary, if the material exhibits several domains.
If the unit cell is chosen to be centered around a filled plaquette as illustrated in Fig.~\ref{Fig:latticeanduc}(a), the positions of the spins within the unit cells are given by
\begin{equation}
\begin{aligned}
\vec{\nu}_1&=\frac{d}{2\sqrt{2}}(1,-1)^T, \qquad \vec{\nu}_2=\tilde h(-1,-1)^T\\
\vec{\nu}_3&=\frac{d}{2\sqrt{2}}(-1,1)^T, \qquad \vec{\nu}_4=\tilde h (1,1)^T .
\end{aligned}
\end{equation}
For the empty plaquettes chosen in Fig.~\ref{Fig:latticeanduc}(b) the position vectors can be written as
\begin{equation}
\begin{aligned}
\vec{\nu}_1&=\frac{a}{2}(1,0)^T +\tilde h(-1,-1)^T, \qquad \vec{\nu}_2=-\frac{a}{2}(1,0)^T + \frac{d}{2\sqrt{2}}(1,-1)^T\\
\vec{\nu}_3&=-\frac{a}{2}(1,0)^T + \tilde h(1,1)^T, \qquad \vec{\nu}_4=\frac{a}{2}(1,0)^T + \frac{d}{2\sqrt{2}}(-1,1)^T.
\end{aligned}
\end{equation}
Similar to the derivation of the excitation energies we use pCUTs to calculate the dynamic structure factors order by order~\cite{0305-4470-36-29-302}. The unperturbed Hamiltonians are the same as before. As for the effective Hamiltonian, we concentrate on the one-QP contributions to the dynamic structure factor in the EPP and FPP/Haldane phase.

For the EPP, the application of the observable in the unperturbed case leads to the creation or annihilation of a single type of triplet on the very same plaquette at $\vec{p}$ in the one-QP sector. One can therefore express the observable as
\begin{align}
\mathcal{O}_{\vec{p},\vec{\nu}} = a_{\text{empty}}^{\vec{\nu}} \left( t_{\vec{p}}^\dagger + t_{\vec{p}}^{\phantom{\dagger}}\right) + \ldots
\end{align}
using the creation and annihilation operators $t_{\vec{p}}^\dagger$ and $t_{\vec{p}}^{\phantom{\dagger}}$. Here $a^\nu$ are the one-triplon amplitudes at $\lambda=0$ and $\ldots$ represents all other QP-processes. 

With the pCUT this process is less confined in space and the area which gets affected by the application of the operator grows with the order of the perturbation theory.
The effective observable in the one-triplon sector reads
\begin{align}
\mathcal{O}_{\vec{p},\vec{\nu}}^{\text{eff},1\text{QP}} = U^\dagger \mathcal{O}_{\vec{p},\vec{\nu}} U \biggr\rvert_{1\text{QP}} = \sum\limits_{\vec{\delta}} a_{\vec{\delta},\text{empty}}^{\vec{\nu}} \left( t_{\vec{p}+\vec{\delta}}^\dagger + t_{\vec{p}+\vec{\delta}}^{\phantom{\dagger}} \right),
\end{align}
where the index $\vec{\delta}$ runs over all plaquettes in infinite order.
In the present case and in finite orders only a finite number of plaquettes is involved.
For the calculation of the dynamic structure factor
\begin{align}
S^{1\text{QP}}(\vec{k},\omega)=-\frac{1}{\pi} \Im \left( \frac{\left\langle 0 \Big| \mathcal{O}_{\text{eff}}^{1\text{QP}\dagger}(\vec{k}) \mathcal{O}^{1\text{QP}}_{\text{eff}}(\vec{k}) \Big| 0 \right\rangle  }{E_0+H_{\text{eff}}^{1\text{QP}}(\vec{k}) - \omega + \rm{i}\delta} \right),
\end{align}
the effective operator is taken in momentum space
\begin{align}
\mathcal{O}_{\text{eff}}^{1\text{QP}}(\vec{k}) = a_{\text{empty}}(\vec{k})\left(t_{\vec{k}}^\dagger + t_{\vec{k}}^{\phantom{\dagger}} \right), \qquad \text{with} \qquad a_\text{empty}(\vec{k}) = \sum\limits_{\vec{\nu}} \sum\limits_{\vec{\delta}} \text{e}^{\textrm{i}\vec{k}\left(\vec{\nu}-\vec{\delta}\right)} a_{\vec{\delta},\text{empty}}^{\vec{\nu}}.
\end{align}
The operators $t_{\vec{k}}^\dagger$ and $t_{\vec{k}}^{\phantom{\dagger}}$ create and annihilate a triplon with momentum $\vec{k}$, respectively.
Finally, the one-triplon part of the dynamic structure factor reduces to
\begin{align}
S^{1\text{QP}}(\vec{k},\omega)=3|a_{\text{empty}}(\vec{k})|^2\delta(\omega(\vec{k})-\omega).
\end{align}
The full information on the intensity is therefore given by $3|a_{\text{empty}}(\vec{k})|^2$, where the factor $3$ accounts for the three spin components $S^\alpha$ with $\alpha\in\{x,y,z\}$ in Eq.~\eqref{eq::dsf}. This quantity is plotted in Fig.~3b in the main body of the manuscript. In order zero of perturbation theory one finds
\begin{align}
a_{\text{empty}}(\vec{k}) = 0.816497 (\cos(0.357615 k_x - 1.21318 k_y) - \cos(1.21188 k_x + 0.358915 k_y)) ,
\end{align}
which is identical to the calculation on a single filled plaquette~\cite{Zayed_plaquette_16}. Note, that here we use the dependency of the momentum, whereas for the comparison with the experiment~\cite{Zayed_plaquette_16} the number of reciprocal lattice units is needed.

For the FPP, one has to perform the same kind of calculation for the one-QP contribution to the dynamic structure factor. However, in this case, two-triplons $\ket{t_1}$ and $\ket{t_2}$ are present in the one-QP sector and the apparent effective Hamiltonian is given by a 2$\times$2 matrix, which needs to be diagonalized to get the proper energy excitations $\ket{t_{\tilde a}}$ and $\ket{t_{\tilde b}}$.
The eigenvectors are denoted by
\begin{equation}
\begin{aligned}
\ket{t_{\tilde a}} &= \tilde a_1 \ket{t_1} + \tilde a_2 \ket{t_2},\\
\ket{t_{\tilde b}} &= \tilde b_1 \ket{t_1} + \tilde b_2 \ket{t_2}.
\end{aligned}
\end{equation}
For the dynamic structure factor of these excitations it is therefore necessary to study the linear combination of contributions
\begin{equation}
\begin{aligned}
S^{\tilde a}(\vec{k},\omega) &= 3\left(\tilde a_1 |a(\vec{k})|^2 + \tilde a_2 |b(\vec{k})|^2 \right) \delta(\omega_{\tilde a}(\vec{k})-\omega),\\
S^{\tilde b}(\vec{k},\omega) &= 3\left(\tilde b_1 |a(\vec{k})|^2 + \tilde b_2 |b(\vec{k})|^2 \right) \delta(\omega_{\tilde b}(\vec{k})-\omega).
\end{aligned}
\end{equation}
The full information on the intensity of each mode is again given by the prefactor of the $\delta$-function taking into account the three spin components $S^\alpha$ with $\alpha\in\{x,y,z\}$ in Eq.~\eqref{eq::dsf}. These quantities are plotted in Fig.~3e in the main body of the manuscript. In lowest-order perturbation theory it is
\begin{align}
a_{\text{full}}(\vec{k}) = 0.816497 \cos(0.358915 (k_x - k_y)) -  0.816497 \cos(1.21318 (k_x + k_y)) .
\end{align}

The matrix elements of the observable are identical for a single empty and filled four-site plaquette.
Therefore, at $\lambda = 0$ only the Fourier-transformation leads to a difference in the dynamic structure factor.
We perform pCUT calculations for both phases using a LCE and reached order five in $\lambda$. Pad\'e extrapolations with the exponents [2,3] and [3,2] are used.

\section{Asymmetric orthogonal-dimer chain}
\begin{figure}[t]
\centering
\includegraphics[width=0.75\columnwidth]{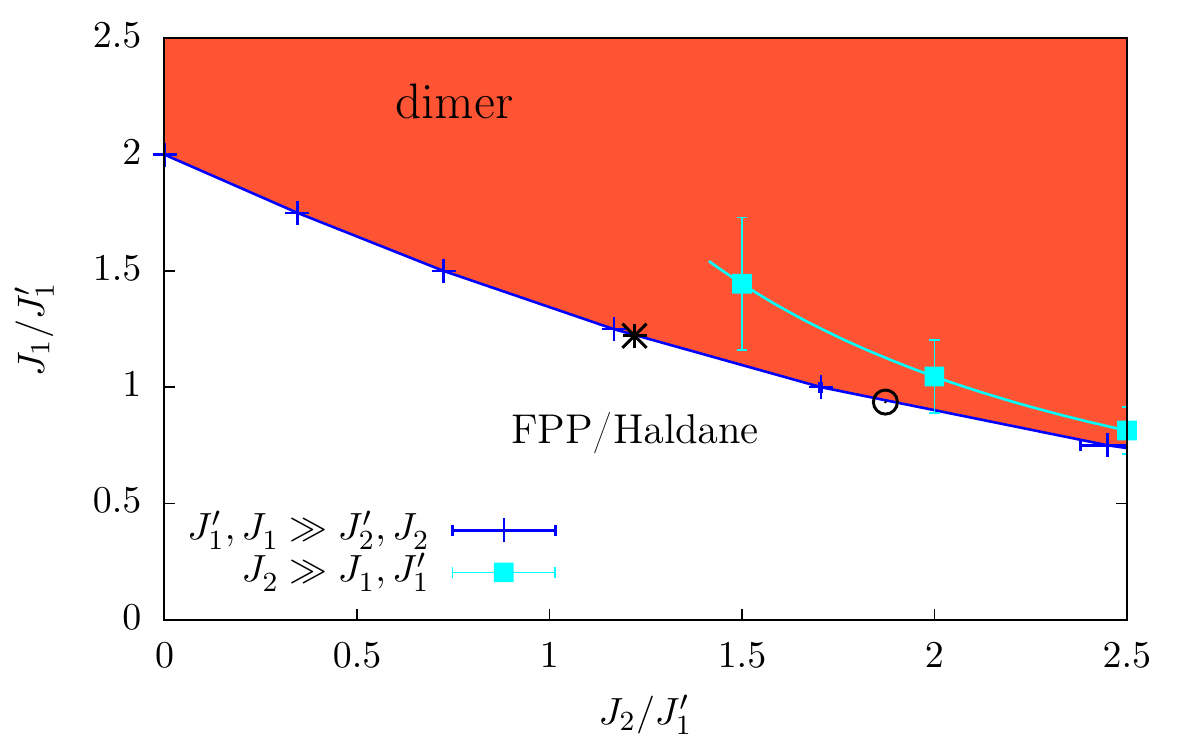}
\caption{The phase diagram of the orthogonal-dimer chain with distinct dimer couplings $J_1$ and $J_2$.
The background color shows results derived by SEs around the limit $J_1',J_1 \gg J_2$.
The red(white) area represents where the dimer singlet(FPP/Haldane) phase is present.
The phase boundary found from the limit $J_2 \gg J_1,J_1'$ and $J_1 \gg J_1'$ is illustrated in cyan.
For comparison two phase transition points from previous works are shown: i) as a black empty circle from~\cite{0953-8984-10-16-015} and ii) as a black star from~\cite{PhysRevB.62.5558}.
}
\label{Fig:phase_diagram_orth_dimer_chain}
\end{figure}
In the limit $J_2'=0$ the distorted Shastry-Sutherland model reduces to decoupled orthogonal-dimer spin chains. This quasi one-dimensional model is very well suited to understand some of the main features also present in the distorted Shastry-Sutherland model.
The phase diagram obtained with SE is shown in Fig.~\ref{Fig:phase_diagram_orth_dimer_chain}.
It exhibits two phases both adiabatically connected to the case of completely decoupled filled plaquettes $J_2=0$:
For weak intra-dimer couplings $J_1$ the ground state is determined by singlets on full plaquettes, where the total spin between both spins on both diagonals is one.
This singlet phase is identical to the Haldane phase.
At $J_1=2$ the ground state changes towards a state which has a total spin zero between the spins on the diagonals, and hence the spins connected by the $J_1$ coupling form a singlet.
This is the exact dimer singlet phase.

In the limit $J_1',J_1 \gg J_2$ at $\lambda = 0$ the decoupled filled plaquettes are present.
In this case the ground state of the unperturbed Hamiltonian is non-degenerate and the ground-state energy follows directly from the SE, which we performed up to order eight in $J_2/J_1'$ and $\Delta J_1/J_1'$. We take the average value of the Pad\'e extrapolations with the exponents [3,4], [4,3] and [4,4] and compare with the dimer singlet energies.
The phase diagram is represented in Fig.~\ref{Fig:phase_diagram_orth_dimer_chain} by the background color (red for the dimer singlet phase, white for the FPP/Haldane phase).

Another limit which can be investigated with SE is $J_2 \gg J_1,J_1'$ and $J_1 \gg J_1'$.
It was previously studied for the asymmetric orthogonal-dimer chain up to second order~\cite{IVANOV1997308} and for an extended Shastry-Sutherland model with distinct dimer couplings $J_1\neq J_2$ in third-order perturbation theory~\cite{PhysRevB.83.140414}.
The ground state of the unperturbed system is degenerate and consists of the manifold of states with a singlet on the $J_2$ bonds and isolated intermediate spins on the (vanishing) $J_1$ bonds.
In third order in $J_1'/J_2$ the effective model
is given by an effective frustrated Heisenberg ladder with rung couplings $J_R$, leg couplings $J_L$, and diagonal couplings between opposite sites of neighboring rungs $J_{\times}$, with $J_L=J_{\times}$.
This effective model is identical for the Shastry-Sutherland model up to order three. Therefore, the FPP in the two-dimensional model is of purely one-dimensional nature for large $J_2$.
In fourth-order perturbation theory additional effective four-spin interactions arise, and the distorted Shastry-Sutherland model is no longer described by a one-dimensional effective model. For the asymmetric orthogonal-dimer chain the effective Hamiltonian in order four is given by
\begin{equation}
\begin{aligned}
&H_{\text{eff}}^{\mathcal{O}(4)}= \epsilon_0 N 
+ J_R \sum_{\scalebox{0.75}{
\begin{tikzpicture}
\def\factor{1.5}
\node (n1) at (\factor*0.75,\factor*0)  [white] {.};
\node (n2) at (\factor*0.75,-\factor*0.43) [white] {.};
\node (n3) at (\factor*0.55,-\factor*0.43) [white] {.};
\node (n4) at (\factor*0.63,\factor*0) [black] {i};
\node (n5) at (\factor*0.63,-\factor*0.43) [black] {j};
 \foreach \from/\to in {n4/n5}
 \draw[black] (\from) -- (\to);
\end{tikzpicture}}}
\vec{S}_i \cdot \vec{S}_j
+ J_L \sum_{\scalebox{0.75}{
\begin{tikzpicture}
\def\factor{1.5}
\node (n1) at (\factor*0.75,\factor*0)  [black] {j};
\node (n4) at (\factor*0.25,\factor*0) [black] {i};
 \foreach \from/\to in {n4/n1}
 \draw[black] (\from) -- (\to);
\end{tikzpicture}}}
\vec{S}_i \cdot \vec{S}_j
+ J_{\times} \sum_{\scalebox{0.75}{
\begin{tikzpicture}
\def\factor{1.5}
\node (n1) at (\factor*0.75,\factor*0)  [black] {j};
\node (n2) at (\factor*0.75,-\factor*0.43) [black] {k};
\node (n4) at (\factor*0.25,\factor*0) [black] {i};
\node (n5) at (\factor*0.25,-\factor*0.43) [black] {l};
 \foreach \from/\to in {n1/n5, n2/n4}
 \draw[black] (\from) -- (\to);
\end{tikzpicture}}}
\vec{S}_i \cdot \vec{S}_k + \vec{S}_j \cdot \vec{S}_l\\
&+ J_K^{R} \sum_{\scalebox{0.75}{
\begin{tikzpicture}
\def\factor{1.5}
\node (n1) at (\factor*0.75,\factor*0)  [black] {j};
\node (n2) at (\factor*0.75,-\factor*0.43) [black] {k};
\node (n4) at (\factor*0.25,\factor*0) [black] {i};
\node (n5) at (\factor*0.25,-\factor*0.43) [black] {l};
 \foreach \from/\to in {n4/n5, n2/n1}
 \draw[black] (\from) -- (\to);
\end{tikzpicture}}}
(\vec{S}_i \cdot \vec{S}_l)(\vec{S}_j \cdot \vec{S}_k)
+ J_K^{L} \sum_{\scalebox{0.75}{
\begin{tikzpicture}
\def\factor{1.5}
\node (n1) at (\factor*0.75,\factor*0)  [black] {j};
\node (n2) at (\factor*0.75,-\factor*0.43) [black] {k};
\node (n4) at (\factor*0.25,\factor*0) [black] {i};
\node (n5) at (\factor*0.25,-\factor*0.43) [black] {l};
 \foreach \from/\to in {n4/n1, n2/n5}
 \draw[black] (\from) -- (\to);
\end{tikzpicture}}}
(\vec{S}_i \cdot \vec{S}_j)(\vec{S}_l \cdot \vec{S}_k)
+ J_K^{\times} \sum_{\scalebox{0.75}{
\begin{tikzpicture}
\def\factor{1.5}
\node (n1) at (\factor*0.75,\factor*0)  [black] {j};
\node (n2) at (\factor*0.75,-\factor*0.43) [black] {k};
\node (n4) at (\factor*0.25,\factor*0) [black] {i};
\node (n5) at (\factor*0.25,-\factor*0.43) [black] {l};
 \foreach \from/\to in {n2/n4, n1/n5}
 \draw[black] (\from) -- (\to);
\end{tikzpicture}}}
(\vec{S}_i \cdot \vec{S}_k)(\vec{S}_l \cdot \vec{S}_j),
\end{aligned}
\end{equation}
with the effective coupling parameters
\begin{equation}
\begin{aligned}
&J_R=J_1 -\frac{J'^2}{J_2} -\frac{1}{2} \frac{J'^3}{J_2^2} + \frac{5}{8} \frac{J'^4}{J_2^3}, \quad
J_L=\frac{1}{2}\frac{J'^2}{J_2}  +  \frac{3}{4} \frac{J'^3}{J_2^2} - \frac{5}{8} \frac{J'^4}{J_2^3},\\
&J_{\times}= J_L, \quad
J_{K}^{R}= -\frac{1}{2} \frac{J'^4}{J_2^3},\quad
J_{K}^{L}= \frac{J'^4}{J_2^3}, \quad
J_{K}^{\times}= J_{K}^{L},
\end{aligned}
\end{equation}
and the constant
\begin{equation}
\begin{aligned}
\epsilon_0= -\frac{1}{4} \frac{J'^2}{J_2} - \frac{7}{8} \frac{J'^3}{J_2^2} + \frac{11}{8} \frac{J'^4}{J_2^3}.
\end{aligned}
\end{equation}
%
For the further analysis the third-order model is particularly useful because it has been studied before~\cite{PhysRevB.93.054408}.
The total spin quantum number on every rung is conserved and in the limit $J_R \gg J_L$, or $J_2 \gg J'^2/J_1$, the system exhibits a ground state with singlets on every rung.
At $J_R/J_L \approx 1.4$ a first-order phase transition takes place to a state where all rungs are occupied by triplets. This state corresponds to a spin-$1$ chain and therefore is associated with the Haldane phase.
In terms of the coupling constants of the asymmetric orthogonal-dimer chain the phase transition is at $J_1\big|_{\text{cr}} \simeq 1.7 \frac{J'^2}{J_2} + 1.55 \frac{J'^3}{J_2^2} + 1.5 \frac{J'^4}{J_2^3}$ (the fourth-order term is not exact due to the additional four-spin interactions).
This phase transition is included as a cyan line in the phase diagram in Fig.~\ref{Fig:phase_diagram_orth_dimer_chain} by the average of the bare second-, third-, and fourth-order result.
The SEs from both limits can be seen to yield similar results.
Additionally, a couple of phase transition points from the literature are included for comparison in the phase diagram in Fig.~\ref{Fig:phase_diagram_orth_dimer_chain}. In the symmetric case $J_1 = J_2$ our SE results agree very well with the value $J/J'|_{\text{cr}} = 1.22100$ by Koga \textit{et al}~\cite{PhysRevB.62.5558}. Along the line $J_2 = 2J_1$ exact diagonalization by Richter \textit{et al} revealed another transition point~\cite{IVANOV1997308}, which also matches our findings.

At last we give some information concerning the magnetic dispersion of the asymmetric-orthogonal dimer chain.
From the limit $J_2 \gg J_1,J_1'$ and $J_1 \gg J_1'$ it is clear that the Haldane phase of the spin-$1$ chain exhibits a low-lying dispersive excitation. The minimum gives the Haldane gap $\Delta_{\text{H}}=0.41$ at $k=\pi$~\cite{PhysRevB.48.3844, PhysRevB.50.3037}. This mode decays at small momenta, due to a continuum~\cite{PhysRevB.48.3844}. In terms of the frustrated ladder the energy gap is $\Delta_{\text{H}} = 0.41 J_L$.
%
Another excitation is given by a rung singlet and is therefore completely localized. The excitation energy of this state is linked to the energy difference between a spin-1 chain with periodic and with open boundary conditions.
It has been determined to be $1.21 J_L$~\cite{PhysRevB.93.054408}.
In terms of the frustrated ladder with an interaction on the bond of the flipped triplet we need to subtract the energy gained by the local singlet and find $\Delta_{\text{f}} = 1.21 J_L - J_R$.

\begin{figure}[t]
\centering
\includegraphics[width=0.75\columnwidth]{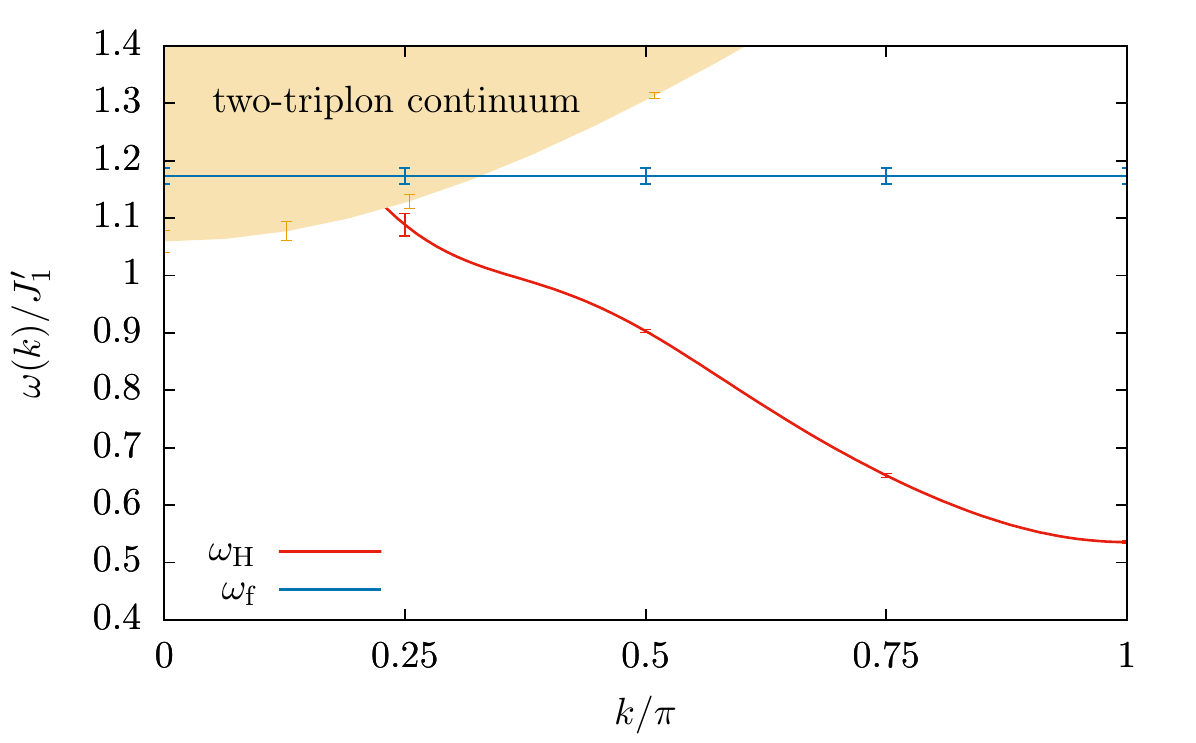}
\caption{Magnetic excitations of the asymmetric orthogonal-dimer chain in the FPP at coupling values $J_1=0.5$ and $J_2=1.2$. This can be understood from the limit $J_2 \gg J_1,J_1'$. The excitation spectrum resembles the one of the Haldane phase on a spin-1 chain.}
\label{Fig:excitations1_orth_dimer_chain}
\end{figure}

Both low-energy excitations together with the continuum of two Haldane triplons from the limit $J_1',J_1 \gg J_2$ at $\lambda = 0$ are depicted for the parameters $J_1=0.5$ and $J_2=1.2$ in Fig.~\ref{Fig:excitations1_orth_dimer_chain}. The Haldane gap at $k=\pi$ is present and the corresponding dispersion increases with decreasing momentum. At values around $k\approx0.45 \pi$ nearly a saddle point can be observed, which becomes more pronounced for larger $J_2$ couplings and eventually transforms into a local maximum. For momenta close to $k=0$ the Haldane mode decays into the continuum. All these three features are also known from the spin-1 Heisenberg chain~\cite{PhysRevB.48.3844} and this set of parameters is already somehow close to the Haldane limit with $J_2 \gg J_1', J_1$ and $J_1 \gg J_1'$.
The flat excitation at these parameters lies above the Haldane mode and also overlaps with the continuum. Nevertheless it does not decay since the total spins on the diagonals $J_1$ are conserved quantities.
The overall structure of the excitation spectrum depends qualitatively on the coupling values.


%

\end{document}